\documentclass[final,3p,times,twocolumn, sort&compress]{elsarticle}

\usepackage{lineno,hyperref}
\modulolinenumbers[5]

\usepackage{bbding,enumitem}
\usepackage{pifont}%

\usepackage{amsmath}
\usepackage{amsfonts,color}
\usepackage{amssymb,float}
\usepackage{mathtools}
\usepackage[utf8]{inputenc}
\usepackage{graphicx}
\usepackage{wasysym}
\usepackage{tabularx}
\usepackage{accents}
\usepackage{graphicx}
\usepackage{color, colortbl}
\usepackage{hyperref}
\usepackage{ulem}

\hypersetup{
    colorlinks=true,
    linkcolor=blue,
    filecolor=magenta,      
    citecolor=red
}

\usepackage{subfigure} 
\usepackage{hyperref} 
\usepackage{multirow}

\usepackage{tikz}
\usetikzlibrary{shapes.geometric, arrows}

\tikzstyle{process} = [rectangle, minimum width=2.2cm, minimum height=1cm, text centered, draw=black]
\tikzstyle{decision} = [diamond, aspect=2, draw=black, text centered, inner sep=1pt, minimum height=1.2cm, minimum width=2cm]
\tikzstyle{arrow} = [thick,->,>=stealth]

\definecolor{Gray}{gray}{0.9}
\definecolor{LightCyan}{rgb}{0.88,1,1}

\newcommand{\be}{\begin{equation}}
\newcommand{\ee}{\end{equation}}
\newcommand{\bea}{\begin{eqnarray}}
\newcommand{\eea}{\end{eqnarray}}

\journal{}

\begin{document}

\begin{frontmatter}

\title{Genetic algorithm demystified for cosmological parameter estimation}


\author[a]{Reginald Christian Bernardo\corref{mycorrespondingauthor}}
\cortext[mycorrespondingauthor]{Corresponding author}
\ead{reginald.bernardo@apctp.org}

\author[c]{Yun Chen}
\ead{s0823042@gm.ncue.edu.tw}

\address[a]{Asia Pacific Center for Theoretical Physics, Pohang 37673, Republic of Korea}
\address[c]{Department of Physics, National Changhua University of Education, Changhua 50007, Taiwan}

\begin{abstract}
Genetic algorithm (GA) belong to a class of nature-inspired evolutionary algorithms that leverage concepts from natural selection to perform optimization tasks. In cosmology, the standard method for estimating parameters is the Markov chain Monte Carlo (MCMC) approach, renowned for its reliability in determining cosmological parameters. This paper presents a pedagogical examination of GA as a potential corroborative tool to MCMC for cosmological parameter estimation. Utilizing data sets from cosmic chronometers and supernovae with a curved $\Lambda$CDM model, we explore the impact of GA's key hyperparameters---such as the fitness function, crossover rate, and mutation rate---on the population of cosmological parameters determined by the evolutionary process. We compare the results obtained with GA to those by MCMC, analyzing their effectiveness and viability for cosmological application.
\end{abstract}


\end{frontmatter}




\section{Introduction}
\label{sec:intro}

Genetic algorithm (GA) is a biology-inspired optimization strategy that incorporates elements of natural evolution to identify the fittest solution from a pool of similarly selected individual solutions. As a powerful optimization method, GA is classified as metaheuristic because it does not rely on derivatives to find the optimum. Under certain conditions, it guarantees the best solution, even overcoming challenges posed by local optimality \cite{rudolph1994convergence}. This method has been applied to a wide range of scientific problems, such as high energy physics \cite{Akrami:2009hp} and gravitational wave astronomy \cite{Crowder:2006wh}. GA is known for its ability to find the global optimum and distinguish tiny differences between seemingly similar solutions. It is particularly effective in navigating complex, high-dimensional parameter spaces and multimodal functions \cite{10.7551/mitpress/1090.001.0001, Katoch2021, Mirjalili2019, Katsifarakis2020, Thompson2024}. In cosmology, it was introduced to overcome biases in selecting a cosmological model for inferring the properties of dark energy \cite{Bogdanos:2009ib}. This approach was further developed by \cite{Nesseris:2010ep} and \cite{Nesseris:2012tt}, which promoted GA as an alternative tool for cosmological analysis through uncertainty estimation. An excellent recent introduction to GA for cosmological parameter estimation can be found in \cite{Medel-Esquivel:2023nov} {(Figure 1 of Ref. \cite{Medel-Esquivel:2023nov} teases out GA's exploratory ability).}

On one hand, cosmology is currently a field facing significant challenges, necessitating a reevaluation of its foundational theories and analytical methods due to several persistent tensions in its fundamental parameters \cite{DiValentino:2025sru, DiValentino:2020vhf, DiValentino:2020zio, DiValentino:2020vvd, Schoneberg:2021qvd}. Notable among these is the Hubble tension \cite{DiValentino:2020zio}, which refers to the discrepancy between the Hubble constant values derived from early universe observations (such as the cosmic microwave background \cite{Aghanim:2018eyx}) and those obtained from local universe measurements \cite{Riess:2021jrx}. Another critical issue is the tension in the amplitude of the smoothed matter power spectrum, which affects our understanding of large-scale structure formation \cite{DiValentino:2020vvd}. These discrepancies suggest potential gaps or inaccuracies in our cosmological models and motivate the search for more robust analytical tools. In this context, GA, with its ability to navigate complex, high-dimensional parameter spaces and identify global optima, appears to be a promising approach \cite{10.7551/mitpress/1090.001.0001, Katoch2021, Mirjalili2019, Katsifarakis2020, Thompson2024}. While not intended to replace traditional methods like Markov Chain Monte Carlo (MCMC), GA can complement them by providing alternative solutions and insights.

Our work provides a detailed and pedagogical exploration of the intricacies of GA for cosmological applications \cite{Kammerer:2025dbi, Sui:2024wob, Bartlett:2024jes, Aizpuru:2021vhd, Bonici:2025ltp, Bartlett:2023cyr}, using parameter estimation in the standard $\Lambda$CDM cosmological model as a medium of instruction. We focus specifically on the influences of key GA hyperparameters---fitness function, mutation, and crossover---on the output population of GA. By systematically examining these components, our study aims to clarify the role each one plays in GA and optimization for cosmological parameter estimation. This analysis is crucial for understanding how different settings and configurations can impact the efficiency and accuracy of GA in exploring cosmological parameter spaces. Additionally, our work demonstrates how GA can complement traditional methods, such as MCMC, by offering alternative pathways to finding optimal solutions \cite{DiValentino:2025sru}. Through this comprehensive approach, we aim to provide valuable guidance for researchers looking to apply GA in their cosmological investigations, thereby enhancing the toolkit available for addressing some of the field's most persistent challenges.

{
Understandably, GA is not the only alternative to MCMC. Our readers are encouraged to look into Refs. \cite{Piras:2023aub, Campagne:2023ter, Ruiz-Zapatero:2023hdf, Bonici:2023xjk, Karwal:2024qpt, Balkenhol:2024sbv} for sampling/optimization alternatives such as auto-differentiable method and profile likelihoods. We would also recommend visiting Section 3 (Data analysis in cosmology) of Ref. \cite{DiValentino:2025sru} (CosmoVerse White Paper).
}

{We proceed by first introducing MCMC and GA (Section \ref{sec:genetic_algorithm}), and follow this by a description of the data sets and the cosmological model that we considered for our analysis (Section \ref{sec:data_sets}).}
We present our main results in Section \ref{sec:results_and_discussion}, emphasizing the exploration of GA hyperparameter space to achieve an optimal evolved population, in contrast with only a single best solution. We also touch on the comparison between GA and MCMC (Section \ref{subsec:summary}), setting a baseline for further calibrating GA for cosmological parameter estimation in par with MCMC.  We conclude by discussing potential further applications of GA in cosmology and extensions of this work.

{Our readers are encouraged to freely utilize our python notebook (in the first author's GitHub repository\footnote{\href{https://github.com/reggiebernardo/ga_demystified}{https://github.com/reggiebernardo/ga\_demystified}}) to facilitate their understanding of the paper and GA inferencing.}

\section{MCMC and genetic algorithm}
\label{sec:genetic_algorithm}

We introduce GA in this section through the eyes of MCMC, by first providing an overview of the traditional method, and then following up with a detailed account of GA.

\subsection{MCMC}
\label{subsec:mcmc}

MCMC methods, grounded in Bayesian statistics, are powerful tools for exploring complex parameter spaces through random sampling. By utilizing a Markovian property, where the current state depends only on the immediate preceding state, MCMC facilitates efficient navigation of high-dimensional probability distributions toward a local solution, well approximated by a posterior \cite{Trotta:2008qt, Lewis:2019xzd, 2020arXiv200505290T}.

Given a data set D and parameters $p$ of a model M, Bayes' theorem provides the framework for incrementally improving our estimates of $p$ based on observed data:
\begin{equation}
    P(p | {\rm D}, {\rm M}) = \dfrac{P({\rm D} | p, {\rm M}) P(p | {\rm M})}{P({\rm D} | {\rm M})},
\end{equation}
where $ P(p | {\rm D}, {\rm M}) $ (posterior) is the probability distribution of the parameters $p$ given D and M; $P({\rm D} | p, {\rm M})$ (likelihood) represents the probability of observing D for given $p$ and M; $P(p | {\rm M}) $ (prior) encodes prior knowledge about $p$; and $P({\rm D} | {\rm M})$ (evidence) acts as a
tool to assess the comparative performance of the model with respect to another or a null hypothesis. For cosmological parameter estimation, our focus is on computing the posterior distribution $P(p | {\rm D}, {\rm M})$. {The agreement of two different models M and M' (e.g., a null hypothesis vs a signal) compared to the data can be quantified with the Bayes factor $P({\rm D}|{\rm M})/P({\rm D}|{\rm M}')$ to tell which model better fits. However, for parameter estimation alone, the focus is on the posterior $P(p|{\rm D},{\rm M})$ for a fixed model M and the Bayesian evidence does not need to be computed $P({\rm D}|{\rm M})$.}

Among numerous MCMC algorithms, the Metropolis-Hastings method remains one of the simplest and most widely used. It constructs a Markov chain that converges to the target posterior distribution. The iterative procedure involves: (1) starting at the current position $X(t)$, a new candidate position $Y$ is proposed from a transition distribution $Q(Y; X(t))$; and (2) the candidate position is accepted with probability:
\begin{equation}
    \min \left( 1,\, \frac{P(Y | {\rm D}, {\rm M})}{P(X(t) | {\rm D}, {\rm M})} \, \frac{Q(X(t); Y)}{Q(Y; X(t))} \right).
\end{equation}
If $Y$ is accepted, the chain moves to $Y$, setting $X(t+1) = Y$; otherwise, it remains at $X(t)$, so $X(t+1) = X(t)$. This ensures that the chain samples according to the posterior distribution in the long run.

The choice of the transition distribution $Q(Y; X(t))$ is critical for the efficiency of the Metropolis-Hastings algorithm \cite{10.1214/aoap/1034625254}. A common and practical choice is a multivariate Gaussian distribution centered at $X(t)$, with a covariance matrix carefully tuned for optimal performance. This ensures a balance between exploration and convergence \cite{10.1093/oso/9780198523567.003.0038}. Poorly chosen transition distributions may result in slow convergence or inefficient sampling. Adaptive methods that adjust the covariance matrix dynamically during the sampling process have proven effective in improving performance in high-dimensional spaces \cite{bj/1080222083, 2010arXiv1011.4381V}.

For this work, we consider flat priors and the likelihoods of CC \eqref{eq:like_cc} and SNe \eqref{eq:like_sne} described in Section \ref{sec:data_sets}. We use the public code \texttt{emcee} to perform MCMC \cite{emcee}.

\subsection{Genetic algorithm}
\label{subsec:ga}

GA belongs to the broader class of Evolutionary Algorithms, which optimize populations of solutions rather than focusing solely on individual candidates \cite{DiValentino:2025sru}. This population-based approach has found diverse applications across fields, including high-energy physics and gravitational wave astronomy \cite{Akrami:2009hp, Crowder:2006wh}. In cosmology, GA was first introduced to mitigate biases in model selection, particularly for analyzing dark energy models \cite{Bogdanos:2009ib}. Since then, it has proven effective in a variety of tasks, such as estimating cosmological parameter uncertainties \cite{Nesseris:2010ep, Nesseris:2012tt}, selecting kernel functions in Gaussian process regression \cite{Bernardo:2021mfs}, optimizing neural network architectures \cite{Gomez-Vargas:2022bsm}, enhancing spectroscopic modeling \cite{Bainbridge2017a, Bainbridge2017b, Lee2020AI-VPFIT}, and aiding model selection with information criteria \cite{Webb2021}. We refer readers to \cite{Medel-Esquivel:2023nov, DiValentino:2025sru} for a recent review of GA’s applications in cosmological parameter estimation and reconstruction.

\begin{figure}[h]
    \centering

    \begin{tikzpicture}[scale=0.75, every node/.style={scale=0.75}]
        \draw[pink, thick] (-0.5,-4.5) rectangle (9.5,2);

        \draw[thick] (0,0) rectangle (7,1);
        \node[draw, minimum width=2cm, minimum height=1cm] at (1,0.5) {$H_{0,1}$};
        \node[draw, minimum width=2cm, minimum height=1cm] at (3.5,0.5) {$\Omega_{k_1}$};
        \node[draw, minimum width=2cm, minimum height=1cm] at (6,0.5) {$\Omega_{m_1}$};

        \draw[->, thick, teal] (1,1) -- ++(0,0.5);
        \draw[->, thick, teal] (3.5,1) -- ++(0,0.5);
        \draw[->, thick, teal] (6,1) -- ++(0,0.5);
        \node[teal] at (6,1.7) {gene};

        \draw[thick, orange] (0,-1.5) rectangle (7,-0.5);
        \node[draw, minimum width=2cm, minimum height=1cm] at (1,-1) {$H_{0,2}$};
        \node[draw, minimum width=2cm, minimum height=1cm] at (3.5,-1) {$\Omega_{k_2}$};
        \node[draw, minimum width=2cm, minimum height=1cm] at (6,-1) {$\Omega_{m_2}$};
        \node[orange] at (8.1,-1) {chromosome};

        \draw[thick] (0,-4) rectangle (7,-3);
        \node[draw, minimum width=2cm, minimum height=1cm] at (1,-3.5) {$H_{0,n}$};
        \node[draw, minimum width=2cm, minimum height=1cm] at (3.5,-3.5) {$\Omega_{k_n}$};
        \node[draw, minimum width=2cm, minimum height=1cm] at (6,-3.5) {$\Omega_{m_n}$};

        \node[pink] at (8.5,-4.0) {population};
    \end{tikzpicture}

    \vspace{1em}

    \begin{tikzpicture}[node distance=1.2cm, scale=0.8, every node/.style={scale=0.85}]
        \node (start) [process] {Initial population};
        \node (fitness) [process, below of=start] {\, Fitness function \, };
        \node (decision) [decision, below of=fitness, yshift=-0.2cm] {Termination};
        \node (solution) [right of=decision, xshift=2.5cm] {Solution};
    
        \node (selection) [process, below of=decision, yshift=-1.6cm, xshift=-2cm] {Selection};
        \node (crossover) [process, right of=selection, xshift=4cm] {Crossover};
        \node (mutation) [process, below of=crossover, yshift=-1.4cm] {Mutation};
        \node (nextgen) [process, left of=mutation, xshift=-4cm] {Next generation};
    
        \draw[arrow, gray] (start) -- (fitness);
        \draw[arrow, gray] (fitness) -- (decision);
    
        \draw[arrow, gray] (decision.east) -- ++(1.7,0) node[midway,above] {Yes} -- (solution.west);
    
        \draw[arrow, teal] (decision.south) -- ++(0,-0.6) node[midway,right] {No} -- (selection.north);
    
        \draw[arrow, teal] (selection.east) -- (crossover.west);
        \draw[arrow, teal] (crossover.south) -- (mutation.north);
        \draw[arrow, teal] (mutation.west) -- (nextgen.east);
    
    \end{tikzpicture}

    \caption{[Top] Anatomy of a population in GA and [Bottom] GA's flowchart.}
    \label{fig:GA-flowchart}
\end{figure}

GA simulates the process of natural selection in biological evolution. By employing mechanisms such as inheritance, crossover, mutation, and selection, GA iteratively evolves a population of solutions toward an optimal result (Figure \ref{fig:GA-flowchart}). Unlike methods like MCMC, which excel in local searches, GA performs global searches across the parameter space, making them particularly suitable for high-dimensional and multimodal optimization problems. However, their effectiveness comes at a computational cost, as they are sensitive to parameter settings, and improperly tuned parameters can lead to suboptimal convergence or stagnation in local minima.

{MCMC has of course also been developed to deal with global searches, such as through simulated annealing and nested sampling \cite{Karwal:2024qpt, Feroz:2011bj}. However, in this work, we shall keep to MCMC in its traditional form as described in Section \ref{subsec:mcmc} and as used in \cite{emcee}.}

In comparison to MCMC, GA excels in exploring a broader search space but require careful parameter tuning and greater computational resources. While MCMC relies on localized random walks to converge to a posterior distribution, GA maintains and evolves a population of solutions, increasing the chances of finding global optima. However, this advantage is counterbalanced by challenges such as determining appropriate population sizes, mutation rates, and crossover probabilities, which often require extensive experimentation. Adaptive techniques can partially address these issues, ensuring efficient exploration without compromising convergence speed.

The workflow of a GA involves several stages (Figure \ref{fig:GA-flowchart}), each inspired by biological evolution:
\begin{itemize}[label=, leftmargin=*]
    \item {\it Initialize population---}A population of candidate solutions, or chromosomes, is randomly generated. Each chromosome represents a potential solution, where its genes correspond to parameters like $H_0$, $\Omega_m$, and $\Omega_k$ in cosmological applications.
    \item {\it Fitness calculations---}The fitness function quantifies the quality of each chromosome. In this work, we use statistical measures such as the log-likelihood \eqref{eq:ff1}, an inverse chi-square \eqref{eq:ff2}, and the likelihood \eqref{eq:ff3} to evaluate fitness.
    \item {\it Selection---}In roulette wheel selection, chromosomes are chosen based on their fitness. Higher fitness chromosomes occupy larger sections of the wheel, increasing their chances of being selected for the next generation.
    \item {\it Crossover---}Pairs of selected chromosomes exchange genetic information to produce offspring. We employ a scattered crossover method, which randomly selects crossover points between parent chromosomes to preserve diversity while fostering convergence.
    \item {\it Mutation---}To maintain genetic diversity, random mutations are introduced into the population \footnote{{For the cosmologist, one can view mutation as a random process that in the context of this work changes one of the cosmological parameters; such as $(H_0, \Omega_{m0}, \Omega_{k0}) \rightarrow (H_0, \Omega_{m0}', \Omega_{k0})$ where $\Omega_{m0} \neq \Omega_{m0}'$. The new $\Omega_{m0}'$ is drawn from a probability distribution defined by the GA.}}. Adaptive mutation, where the mutation rate depends on the population’s overall fitness, ensures effective exploration during early generations and stabilizes convergence in later stages.
\end{itemize}

{GA, mathematically, is a global optimization strategy, designed at overcoming the pitfalls such as prior dependencies of classical local search routines \cite{10.7551/mitpress/1090.001.0001, Katoch2021, Mirjalili2019, Katsifarakis2020, Thompson2024}. GA can also be viewed as a population method; where a population ${\bf p}$ is acted upon by a set of probabilistic operators ${\bf O}$, until the maximum number of generations is met or the algorithm reaches termination condition. In this case, the population is classified according to their fitness $f$ and the random operators are selection ${\bf O}_{\rm Sel}$, crossover ${\bf O}_{\rm Cr}$, and mutation ${\bf O}_{\rm Mut}$, each of which are a function of their hyperparameters such as the mutation type and probability \cite{SCHMITT20011, 2011arXiv1105.3538W}. GA mimics natural selection through the action of an operator that can be viewed macroscopically through the population or microscopically via the genetic level in an individual chromosome (Figure \ref{fig:GA-flowchart}). Intrinsically, each operator can be thought of as applying a probability density function to an individual's genetic information in order to keep diversity at the same time looking for possible improvements at the population level. Then, the population produced by the $n$th generation is \cite{10.7551/mitpress/1090.001.0001}
\begin{equation}
{\mathbf p}_{n+1}(f_{n+1})={\mathbf O}_{\rm Mut} {\mathbf O}_{\rm Cr} {\mathbf O}_{\rm Sel} {\mathbf p}_n(f_n) \,,
\end{equation}
and the final evolved population can be written as a successive application of the operators: 
\begin{equation}
{\mathbf p}_N(f_N)=\left({\mathbf O}_{\rm Mut} {\mathbf O}_{\rm Cr} {\mathbf O}_{\rm Sel}\right)^N {\mathbf p}_1(f_1)
\end{equation}
where ${\mathbf p}_1$ is the first (prior) population distribution. The population can be expected to cluster toward the solution in the limit when $N$ becomes infinitely large.}

For this work, we shall play with the fitness function, crossover and mutation rates to assess their impacts on the population. We use the public code \texttt{pyGAD} to perform GA \cite{pygad}.

\section{Cosmology: model and data sets}
\label{sec:data_sets}

We consider background cosmological data from cosmic chronometers (CC) \cite{Moresco:2024wmr, 2010JCAP...02..008S, 2012JCAP...08..006M, 2014RAA....14.1221Z, Moresco:2015cya, Moresco:2016mzx, Ratsimbazafy:2017vga} and supernovae (SNe) \cite{Brout:2021mpj, Brout:2022vxf, Scolnic:2021amr}, described below within the framework of a spatially curved $\Lambda$CDM model.

{\it Cosmic Chronometers---}The CC are best thought of as cosmic standard clocks; due to their inherent capability to give a direct account of the expansion rate of the Universe at late times \cite{Moresco:2024wmr, Moresco:2015cya, Moresco:2016mzx, Moresco:2023zys, Moresco:2018xdr}, $z\lesssim 2$. The main idea is that the expansion rate at a redshift $z_{\rm CC}$ can be estimated through the difference in the ages of redshift-adjacent galaxies, playing on an age-ladder analogous to the usual distance-ladder. Thus, to a good approximation, the expansion rate at a redshift $z_{\rm CC}$ can be written as $H_{\rm CC} \approx -\left(\Delta z_{\rm CC}/\Delta t\right)/\left(1 + z_{\rm CC}\right)$, where $\Delta t$ and $\Delta z_{\rm CC}$ are deduced from the differences in the age and metalicity of temporally-adjacent-passive galaxies. This is a most direct way to observe the cosmic expansion. For our purposes, we consider the CC in the redshift range $0.07 \lesssim z \lesssim 1.97$ that is compiled in \cite{Moresco:2020fbm}.

We describe the cosmology underlying CC as a curved $\Lambda$CDM cosmological model; where the normalized Hubble expansion rate, $E(z) = H(z)/H_0$, is given by
\begin{equation}
\label{eq:normalized_expansion_rate_lcdm}
    E\left(z \right)^2 = \Omega_{m0} \left(1 + z \right)^3 + \Omega_{k0} \left( 1 + z \right)^2 + \left(1 - \Omega_{m0} - \Omega_{k0}\right) \,,
\end{equation}
and $H_0$ is the Hubble constant, $\Omega_{m0}, \Omega_{k0}$ are the energy fractions/density parameters corresponding to nonrelativistic matter and curvature at redshift $z = 0$. The present dark energy fraction is given by $\Omega_{de0}=1-\Omega_{m0}-\Omega_{k0}$, the last term in the parenthesis on the right hand side in \eqref{eq:normalized_expansion_rate_lcdm}. The likelihood of this model compared with the data, $H_{\rm CC}\left(z_{\rm CC}\right) \pm \Delta H_{\rm CC}\left(z_{\rm CC}\right)$ at $z_{\rm CC}$, can be written as
\begin{equation}
\label{eq:like_cc}
\begin{split}
\log {\cal L}_{\rm CC} \propto -\dfrac{1}{2} \sum_{z_{\rm CC}} & \left( H\left(z_{\rm CC}\right) - H_{\rm CC}\left(z_{\rm CC}\right) \right) \\
& \times C_{\rm CC}^{-1} \left( H\left(z_{\rm CC}\right) - H_{\rm CC}\left(z_{\rm CC}\right) \right) \,,
\end{split}
\end{equation}
where $C_{\rm CC}$ is the covariance matrix of correlated CC measurements {used in} \cite{Moresco:2020fbm}.

{\it SuperNovae---}Type Ia SNe are intrinsically very bright objects, historically serving as one of cosmology's most reliable distance indicators \cite{SupernovaCosmologyProject:1997zqe, SupernovaSearchTeam:1998fmf, SupernovaCosmologyProject:1998vns, HST:2000azd, SNLS:2005qlf, Riess:2016jrr, Pan-STARRS1:2017jku}; for this reason, they have been thought of as cosmic standard candles, serving like lampposts in the sky that indicate distances relative to an observer, assuming that the intensity of the source declines according to the inverse square law. Their high brightness and flux particularly facilitate standardization, meaning that by calibrating the flux measurements relative to a host, one can construct a rigorous distance-redshift ladder. This ladder accounts for the Universe's expansion rate through its perceived effect on the brightness of SNe. This distance ladder can be written as
\begin{equation}
\label{eq:distance_modulus_lcdm}
    \mu(z) = 5 \log_{10}\left( d_L\left(z\right) \right) + 25 \,,
\end{equation}
where $\mu(z) = m(z) - M$ is the distance modulus, representing the difference between the apparent $m(z)$ and intrinsic $M$ magnitudes of SNe at a redshift $z$, and $d_L(z)$ is the luminosity distance, estimated through flux measurements, $F = L / (4 \pi d_L^2)$ with $L$ being the luminosity. For this work, we use the Pantheon+ SNe compilation \cite{Brout:2021mpj, Brout:2022vxf, Scolnic:2021amr}, utilizing measurements in the redshift range $0.01 \lesssim z \lesssim 2.3$ to avoid possible bias from galaxy peculiar velocities, among other effects, at very low redshifts $z < 0.01$ \cite{Pasten:2023rpc, Perivolaropoulos:2023iqj}.

As in CC, we attribute the cosmology underlying SNe observations to a curved $\Lambda$CDM model. In this case, the luminosity distance is given by
\begin{equation}
    d_L(z) = \left(1 + z\right) \dfrac{c}{ H_0 \sqrt{ \Omega_{k0}} } \sin \left( H_0 \sqrt{\Omega_{k0}} \int_0^z \dfrac{dz'}{E(z')} \right) \,,
\end{equation}
where $c$ is the speed of light in vacuum and the normalized expansion rate $E(z)$ is given by \eqref{eq:normalized_expansion_rate_lcdm}. We associate this model to the data via the likelihood
\begin{equation}
\label{eq:like_sne}
\begin{split}
    \log {\cal L}_{\rm SNe} \propto -\dfrac{1}{2} \sum_{z_{\rm SNe}} & \left( \mu(z_{\rm SNe}) - \mu_{\rm SNe}(z_{\rm SNe}) \right) \\
    & \times C^{-1}_{\rm SNe} \left( \mu(z_{\rm SNe}) - \mu_{\rm SNe}(z_{\rm SNe}) \right) \,,
\end{split}
\end{equation}
where $C_{\rm SNe}$ is the covariance matrix \cite{Brout:2021mpj, Brout:2022vxf, Scolnic:2021amr}.

\begin{figure}[h!]
    \centering
    \includegraphics[width=0.475\textwidth]{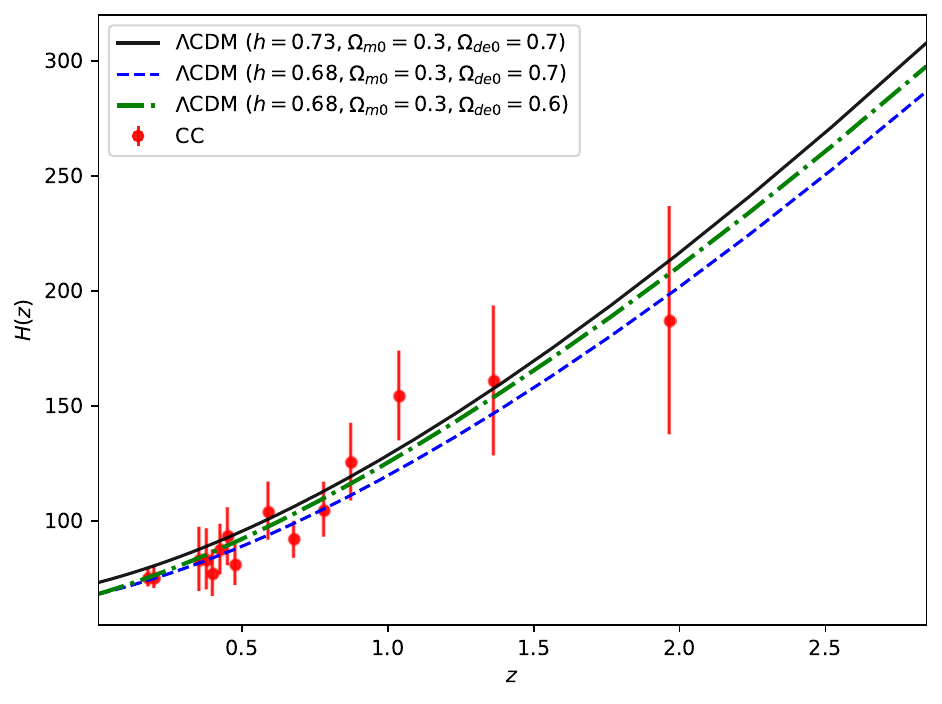}
    \includegraphics[width=0.475\textwidth]{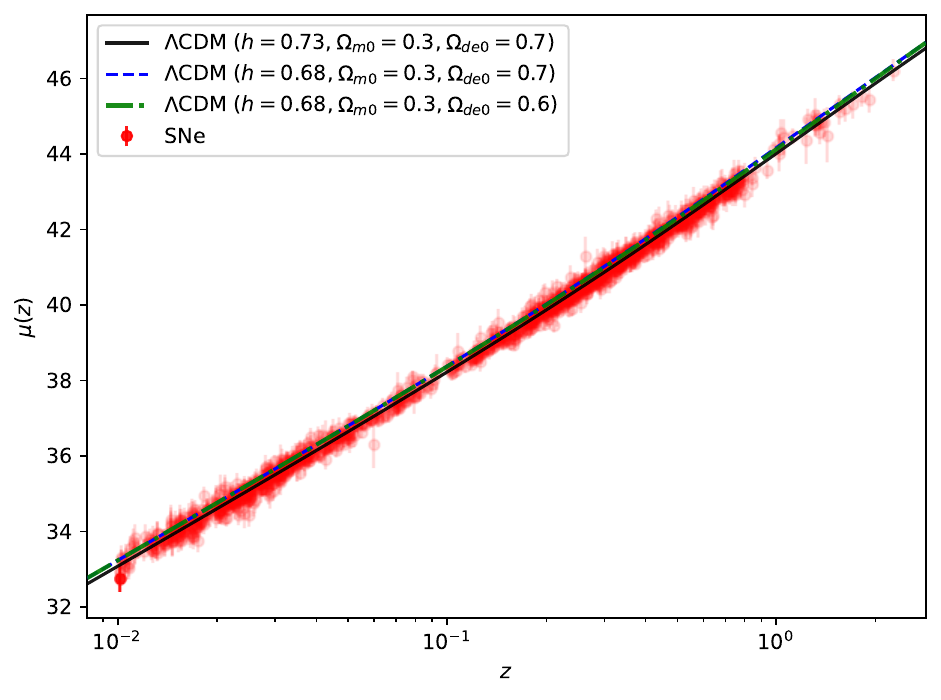}
    \caption{[Left] Expansion rate from CC \cite{Moresco:2020fbm} and [right] distance moduli from SNe \cite{Brout:2021mpj, Brout:2022vxf, Scolnic:2021amr} with curved $\Lambda$CDM predictions \eqref{eq:normalized_expansion_rate_lcdm} and \eqref{eq:distance_modulus_lcdm}. Note that $h=H_0/100$ km s$^{-1}$Mpc$^{-1}$ and $\Omega_{de0}=1-\Omega_{m0}-\Omega_{k0}$.}
    \label{fig:data_and_model}
\end{figure}

Figure \ref{fig:data_and_model} shows the data sets considered in this work together with predictions of the curved $\Lambda$CDM model.

We emphasize that in practice there can be several models that are well suited to fit the same data sets. For this work, we shall not be concerned with model selection and hypothesize that curved $\Lambda$CDM is suitable to describe CC and SNe, such that its parameters can be constrained by MCMC and GA.

\section{Results and Discussion}
\label{sec:results_and_discussion}

In this section, we present our results, playing with the key parameters of GA (fitness function, cross-over, mutation) and studying their influence on the final evolved population, of $\Lambda$CDM cosmological parameters.

\subsection{Fitness function}
\label{subsec:fitnessfunction}

\begin{figure*}[h!]
  \centering
  \includegraphics[width=0.495\textwidth]{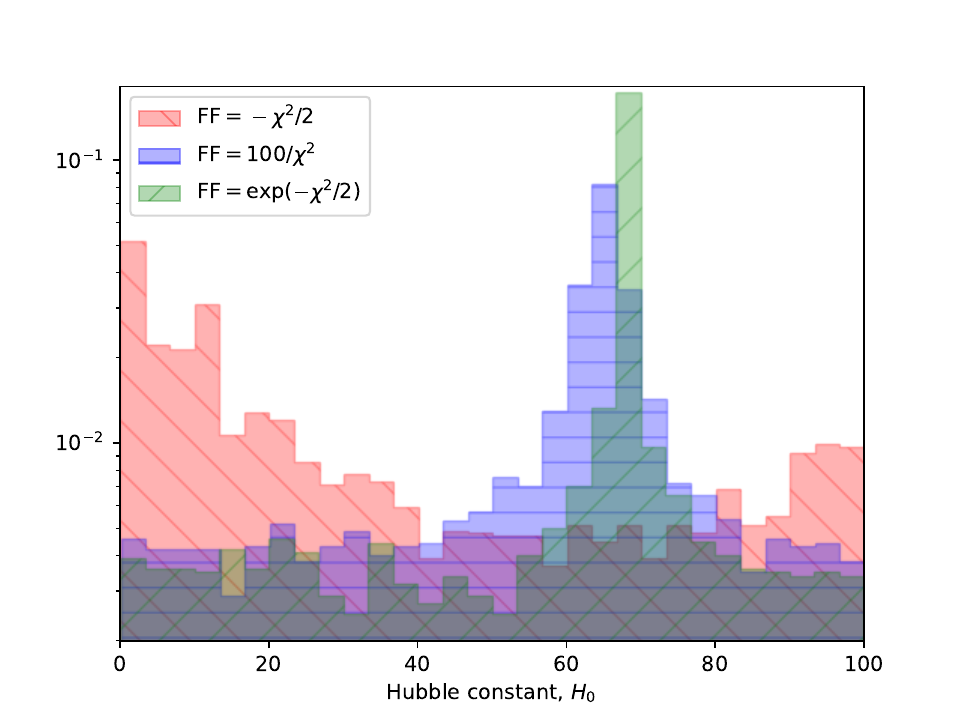}
  \includegraphics[width=0.495\textwidth]{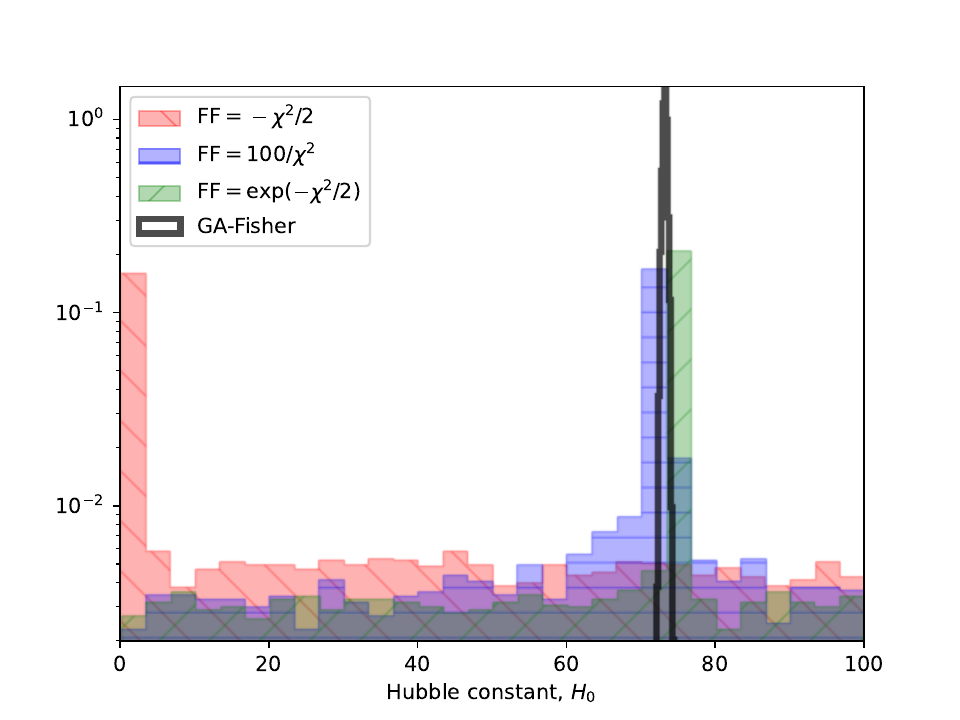}
  \includegraphics[width=0.495\textwidth]{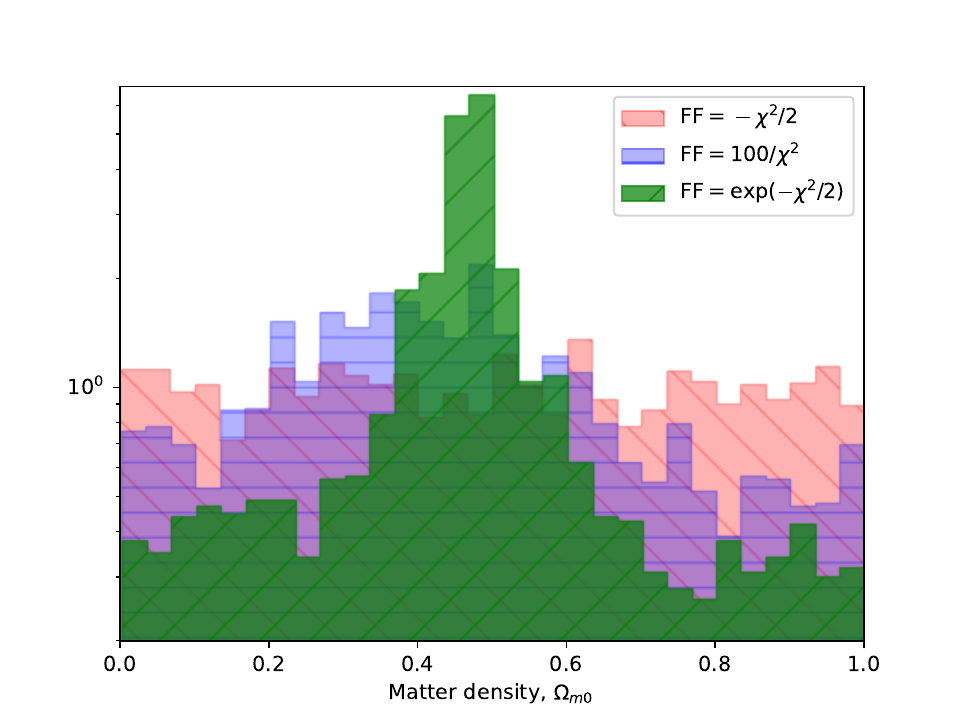}
  \includegraphics[width=0.495\textwidth]{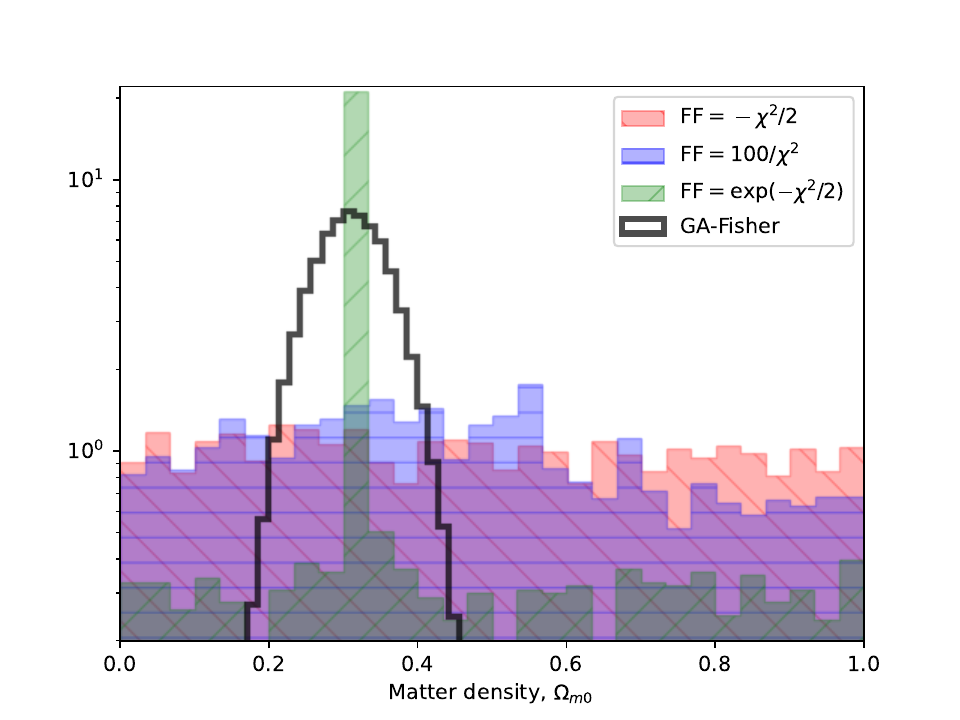}
  \includegraphics[width=0.495\textwidth]{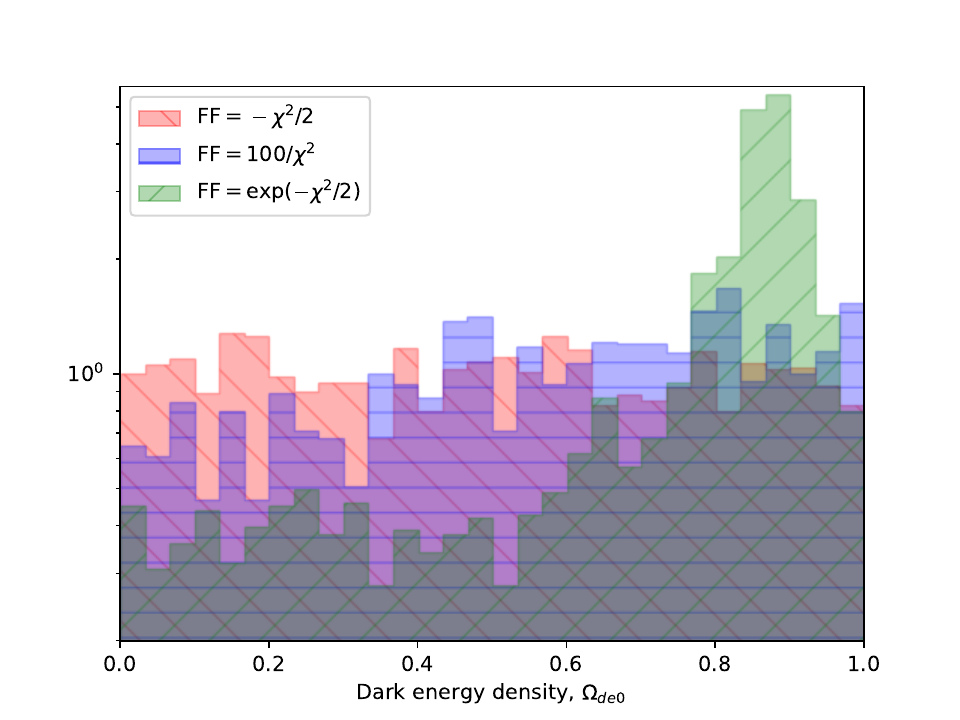}
  \includegraphics[width=0.495\textwidth]{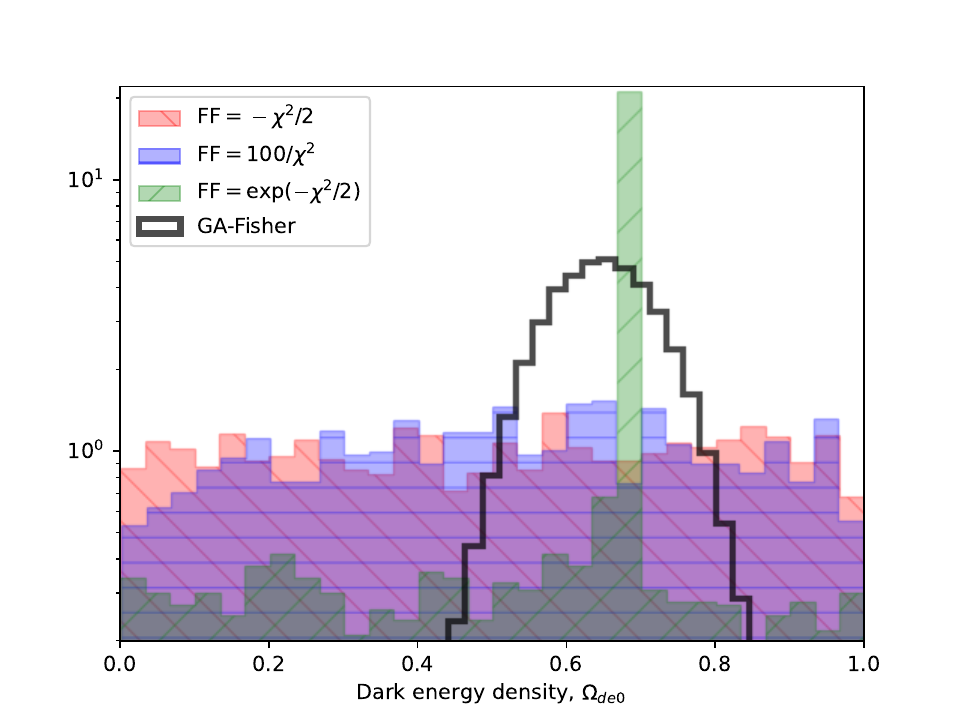}
  \caption{Changing fitness---GA final population distribution for different fitness functions; FF=$-\chi^2/2, 100/\chi^2, \exp(-\chi^2/2)$ with fixed mutation (0.5,0.3) and crossover (50\%) rates. ({L}eft column) results obtained with only CC data{;} (right column) results with CC and SNe. In the right column, the Gaussian (GA-Fisher) corresponds to the GA best solution compounded with a Fisher matrix uncertainty estimate based on the likelihood.}
  \label{fig:changing_fitness}
\end{figure*}

We begin our discussion with GA's most important ingredient, the fitness function. To understand its impact on the parameter estimation, we consider three different functional forms, all depicting a distance measurement between the data and the model prediction; we take
\begin{equation}
\label{eq:ff1}    {\rm FF}_1 = -\chi^2/2 \,,
\end{equation}
\begin{equation}
\label{eq:ff2}    {\rm FF}_2 = 100/\chi^2 \,,
\end{equation}
\begin{equation}
\label{eq:ff3}    {\rm FF}_3 = \exp(-\chi^2/2) \,,
\end{equation}
where $\chi^2 = -2 \log {\cal L}$ and ${\cal L}$ is the likelihood, \eqref{eq:like_cc} for CC and \eqref{eq:like_sne} for SNe. The common denominator between these different forms are that closer distances (between data and model; $\chi^2 \lesssim 1$) give more positive values of the fitness. We feed each fitness function into GA, with an initial uniform prior on the cosmological parameters, and see where the evolution in 100 generations take the population. Throughout, we consider a fixed population size of $3000$, a `roulette wheel' selection with a fixed selection rate of 30\%, a `scattered' crossover with a fixed probability of 50\%, and an adaptive mutation rate, mutating a fraction 30\% of the genes of high quality solutions and a fraction 50\% of the genes of low quality solutions. The evolved population are shown in Figure \ref{fig:changing_fitness} after 100 generations.

{We must say that information theory wise, all these metrics (fitness functions) are the same. However, under a population method such as GA, they would act differently since probabilistic operators (selection, crossover, and mutation) act on the population, discarding and adding new members at every new generation based on fitness.}

Results show that the choice of the functional form of the fitness function has a significant impact to the final evolved population. Focusing on CC results (left column of Figure \ref{fig:changing_fitness}), we realize this in each one of the cosmological parameters, $H_0$, $\Omega_{m0}$, and $\Omega_{de0}$, where each population basically turns out flat with FF$_1$. It is worth noting that we present the histogram frequencies in log-scale in order to highlight the differences between the results of fitness functions (\ref{eq:ff1}-\ref{eq:ff3}), which otherwise would have been more difficult to perceive visually in linear scale. On the other hand, using FF$_2$ gets the final evolved population to be a bit more localized, in the usual way cosmological parameter estimates are viewed. However, the population remains less localized than desired for this purpose. Moving forward to FF$_3$, we find that the last evolved population has turned out quite localized, narrowed to the place where we expect the bulk of the cosmological parameters inferred via MCMC (to be confirmed in Section \ref{subsec:summary}).

The addition of SNe data supports the results obtained with CC data on the importance of picking out a suitable fitness function. Using the same set of fixed GA parameters, we perform the GA with fitness functions (\ref{eq:ff1}-\ref{eq:ff3}). The results after 100 generations are shown in the right column of Figure \ref{fig:changing_fitness}.

{Note that in addition to population distribution, we now also show Fisher matrix uncertainty estimates evaluated at the GA best fit (GA-Fisher) for the CC + SNe cases. The corresponding estimates exceed the priors for the CC only cases. The Fisher matrix ${\mathbf F}$ is evaluated as \cite{Verde:2009tu, Wolz:2012sr, Schafer:2016vyy}
\begin{equation}
    {\mathbf F}_{pp'} = - \langle \partial_p \partial_{p'} \log {\cal L} \rangle \,,
\end{equation}
where ${\cal L}$ is the likelihood, $p$ and $p'$ are model parameters, and $\langle \cdots \rangle$ denotes an expectation value over the data. Then, the uncertainty is estimated using a Gaussian distribution with a covariance matrix ${\mathbf F}$. The Fisher matrix bands are a useful reference point to tell visually that the GA generates highly non-Gaussian parameter distributions. The GA-Fisher approach has been shown reliable in reconstructing cosmological functions \cite{Nesseris:2012tt} and cosmological parameter estimation \cite{Medel-Esquivel:2023nov}.}

This shows that the fitness function FF$_1$ leads to a flat final population (parameter distribution) despite all the operations and evolution in GA; while FF$_2$ leads to kind of the same uniform distribution, except for $H_0$ which somehow turns out localized. This suggests that the addition of the more stringent supernovae data is insufficient to help GA get out of the initial uniform population distribution. However, the fitness function FF$_3$ completely amends this, in all cases resulting to a quite localized distribution of parameters after 100 generations. The best GA solution almost always belongs to this localized set in the final population. The impact of the addition of SNe on the width of the localized set per parameter is also highly notable.

{
Our chosen hyperparameters (population size, selection rate, crossover probability, and mutation fractions) are not optimal or universal choices. Rather, we picked them to illustrate how different settings could affect the population distribution and, consequently, the performance of the algorithm. We aimed to balance computational tractability with sufficient genetic diversity, while ensuring that the effects of crossover and mutation could be visible. These values are well within the range typically employed \cite{pygad} but tuned them slightly so that the qualitative trends we wished to highlight would be visible. Granted, solving a problem with GA comes with determining a suitable set of hyperparameters. The choices given have been shown to work suitably well for the data sets we considered and for cosmological models beyond the ones considered in this work \cite{DiValentino:2025sru, Bernardo:2025pua}.
}

\subsection{Mutation}
\label{subsec:mutation}

\begin{figure*}[h!]
  \centering
  \includegraphics[width=0.495\textwidth]{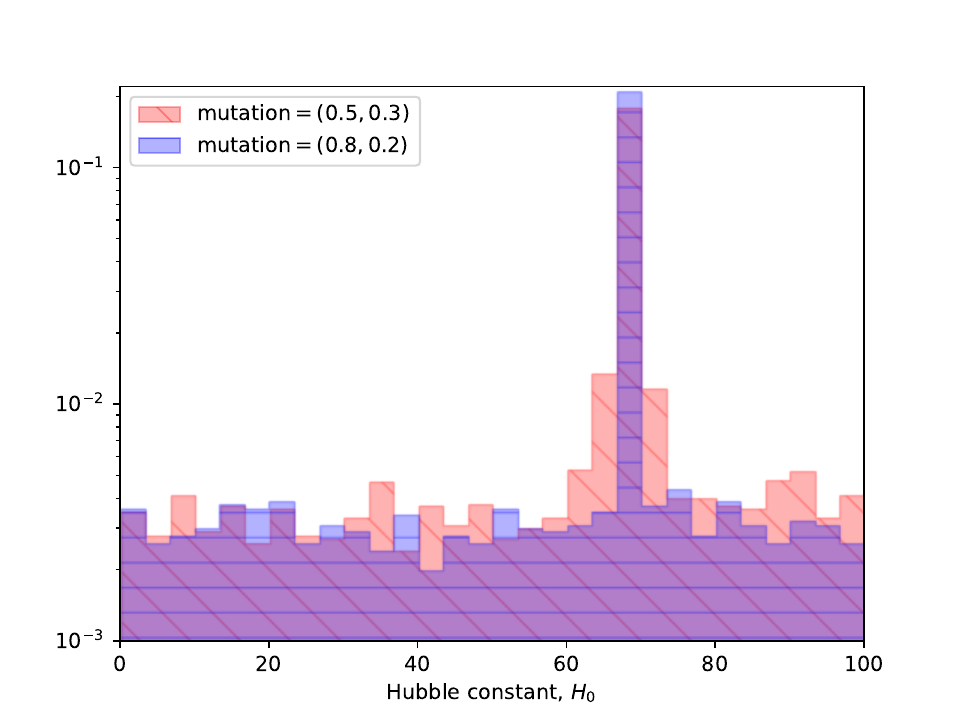}
  \includegraphics[width=0.495\textwidth]{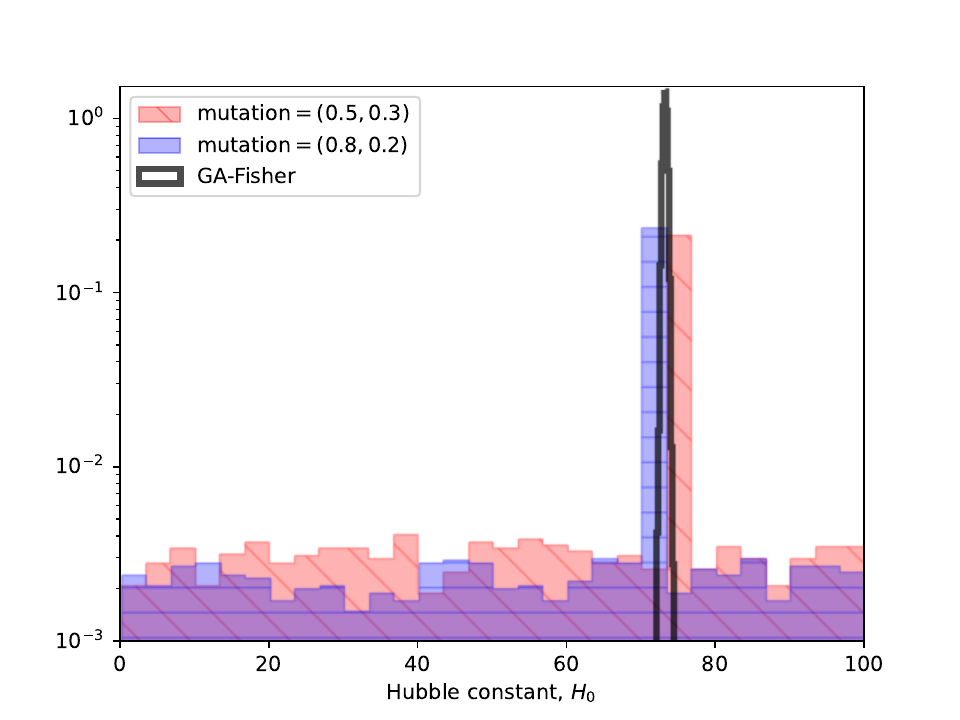}
  \includegraphics[width=0.495\textwidth]{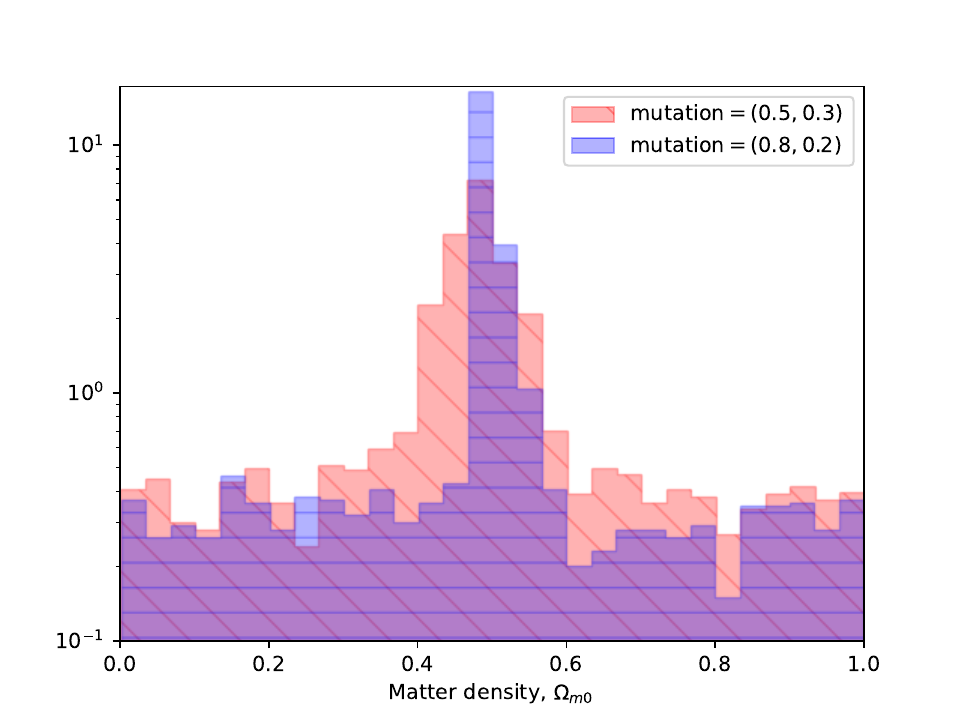}
  \includegraphics[width=0.495\textwidth]{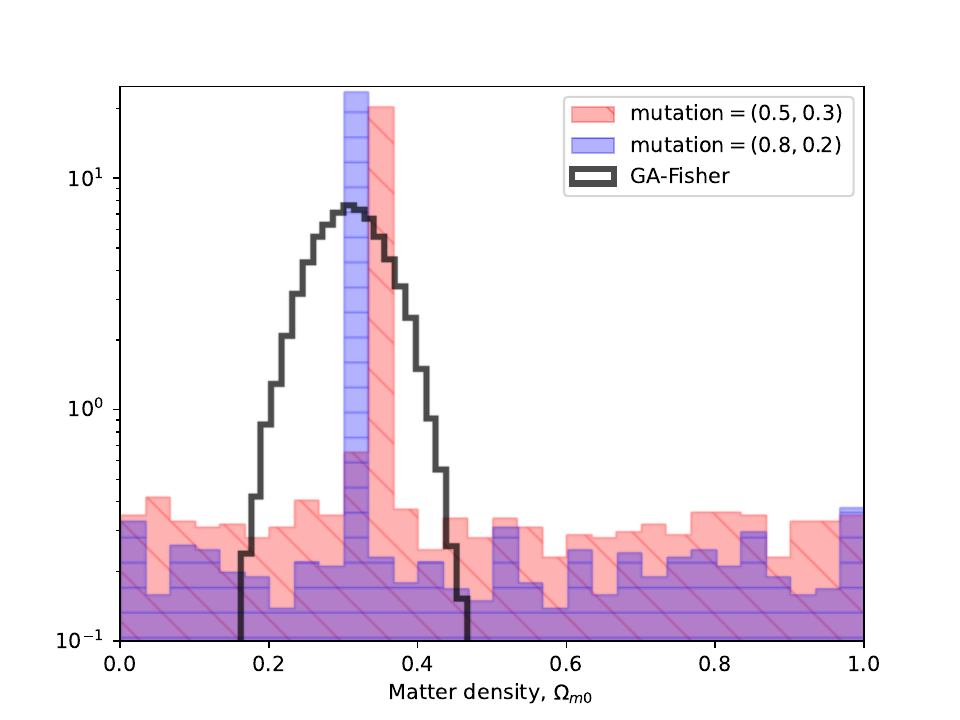}
  \includegraphics[width=0.495\textwidth]{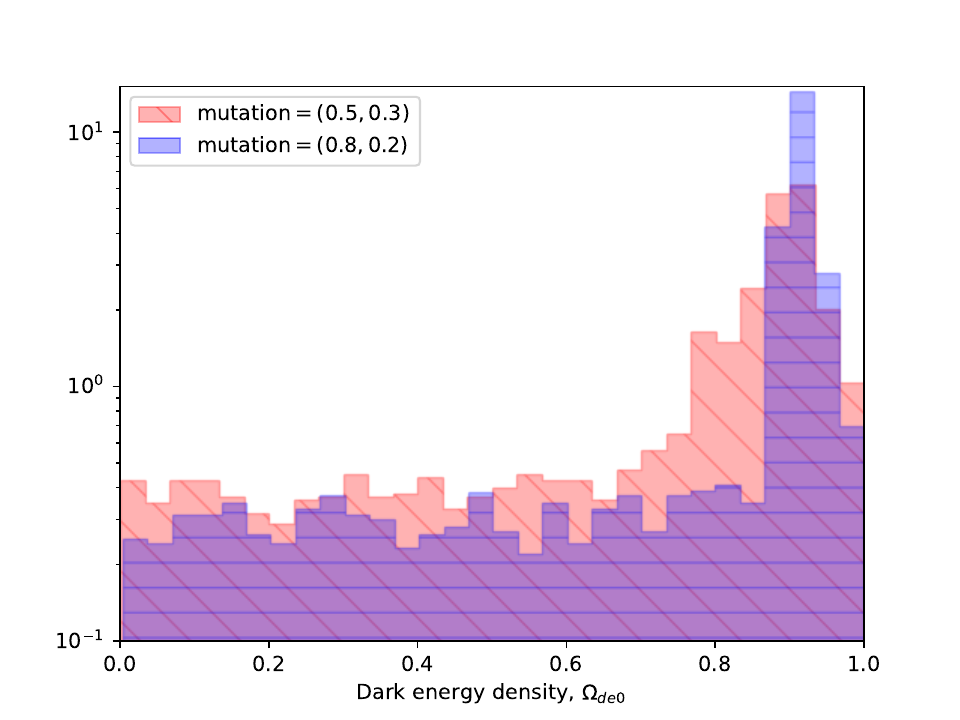}
  \includegraphics[width=0.495\textwidth]{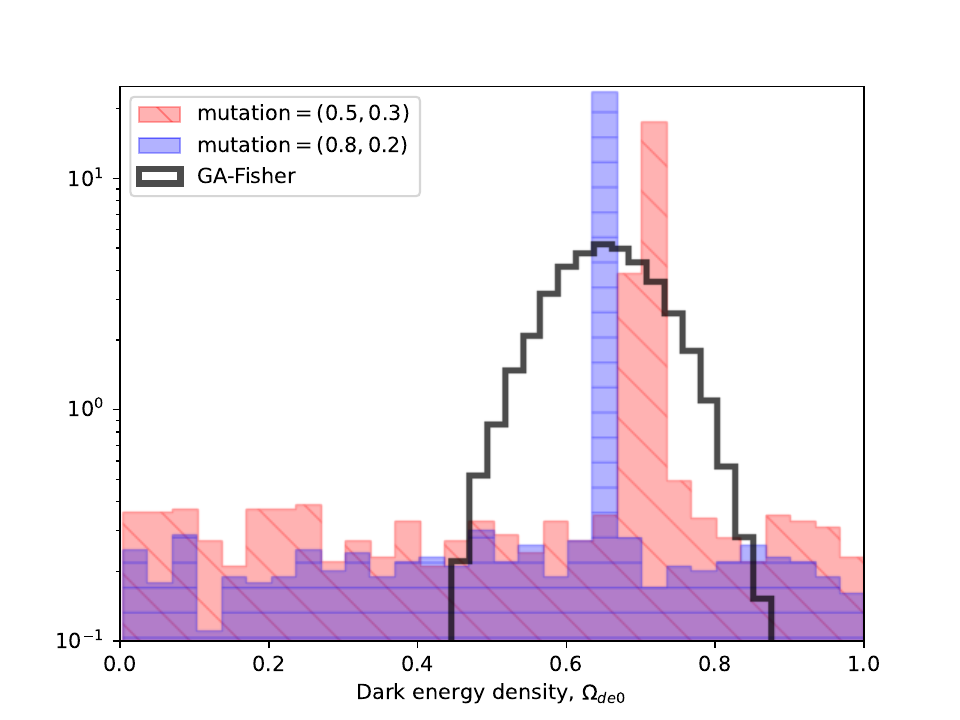}
  \caption{Changing mutation---GA final population distribution for different mutation rates; mutation=$(0.5, 0.3), (0.8, 0.2)$ with fixed fitness function $\exp(-\chi^2/2)$ and crossover rate (50\%). {(Left column) results obtained with only CC data; (right column) results with CC and SNe.} In the right column, the Gaussian (GA-Fisher) corresponds to the GA best solution compounded with a Fisher matrix uncertainty estimate based on the likelihood.}
  \label{fig:changing_mutation}
\end{figure*}

We next explore the impact of mutation (or rather the mutation rate) on the final evolved population in GA. Following the lessons of the previous section, we fix the fitness function to be FF$_3$ \eqref{eq:ff3}, and consider two choices for an adaptive mutation rate of $m_1=(50\%, 30\%)$ and $m_2=(80\%, 30\%)$. The first choice $m_1$ means that half of the chromosomes of a gene are going to be tweaked by mutation for low quality solutions; while 30\% of the chromosomes will be altered for high quality solutions. The same interpretation extends to our second choice $m_2$. Ideally we want the first percentage in the input tuple to be higher such that low quality solutions mutate more. In addition to a fixed fitness function, we consider a fixed population size of 3000, a roulette wheel selection with a fixed selection rate of 30\%, and a scattered crossover with a fixed probability of 50\%. The evolved population are shown in Figure \ref{fig:changing_mutation} after 100 generations. 

Our results suggest that higher mutation rates also lead to more localized populations after evolution. To get to this, we start with the CC results (left column of Figure \ref{fig:changing_mutation}). In this case, we find that the populations resulted from the higher mutation $m_2$ are always a tad narrower compared with the populations outputed by $m_1$. Nonetheless, the peaks of the parameter space histograms resulted from both mutation rates coincide. The same conclusion can be drawn with the addition of SNe data (right column of Figure \ref{fig:changing_mutation}). The final populations per parameter have localized as expected due to the addition of a stringent data set. Consequently, the difference between the widths of the localized set of parameters in the final population has turned out to be less perceivable. In principle, there may still be a small difference. However, we expect that such minute difference will be completely hidden away by uncertainty, which we have yet to return to (in Section \ref{subsec:summary}).

Our results support that higher mutations can have help in producing a localized set of parameters in the final evolved population.

\subsection{Crossover}
\label{subsec:crossover}

\begin{figure*}[h!]
  \centering
  \includegraphics[width=0.495\textwidth]{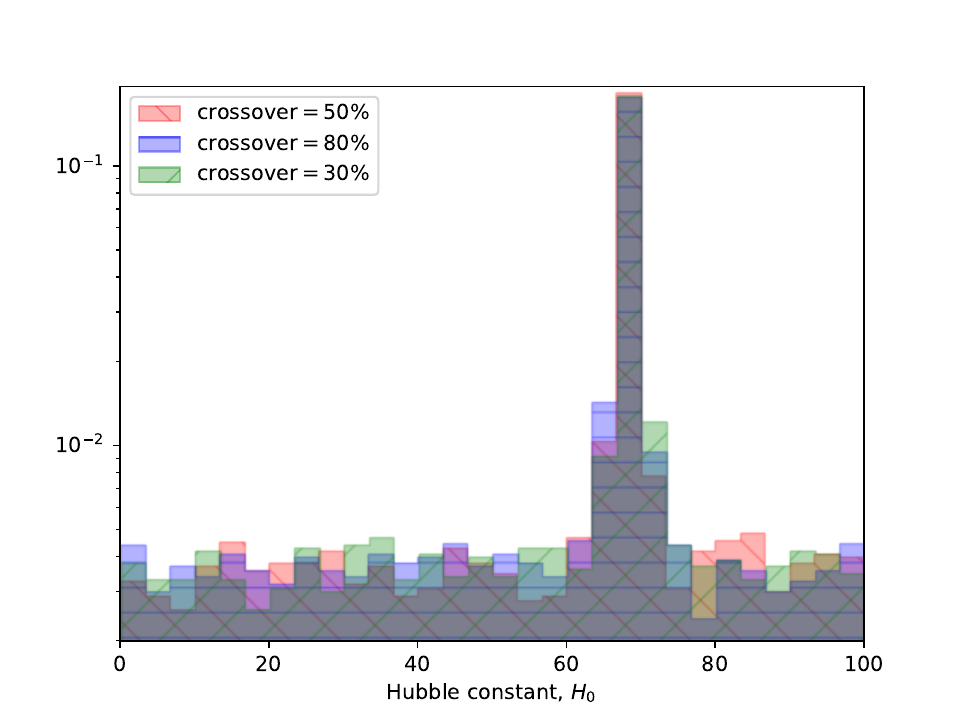}
  \includegraphics[width=0.495\textwidth]{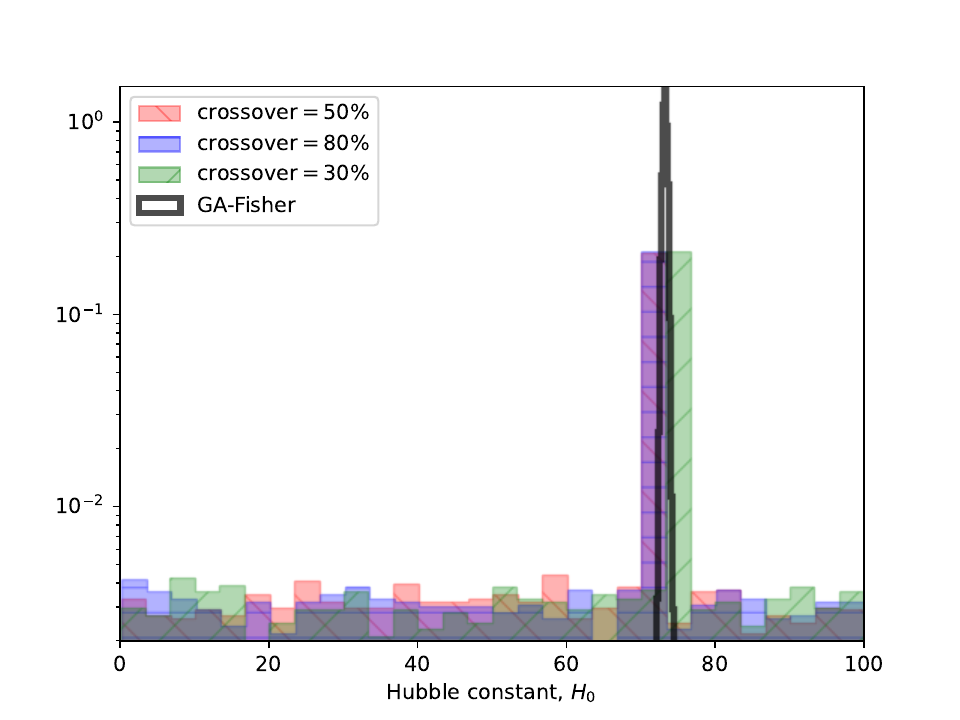}
  \includegraphics[width=0.495\textwidth]{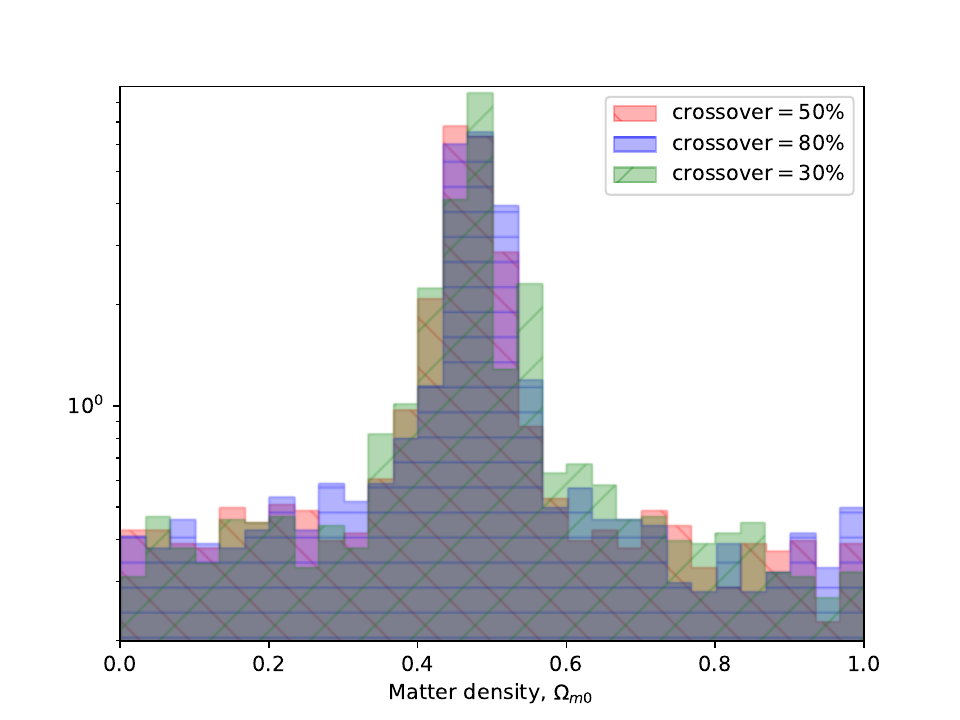}
  \includegraphics[width=0.495\textwidth]{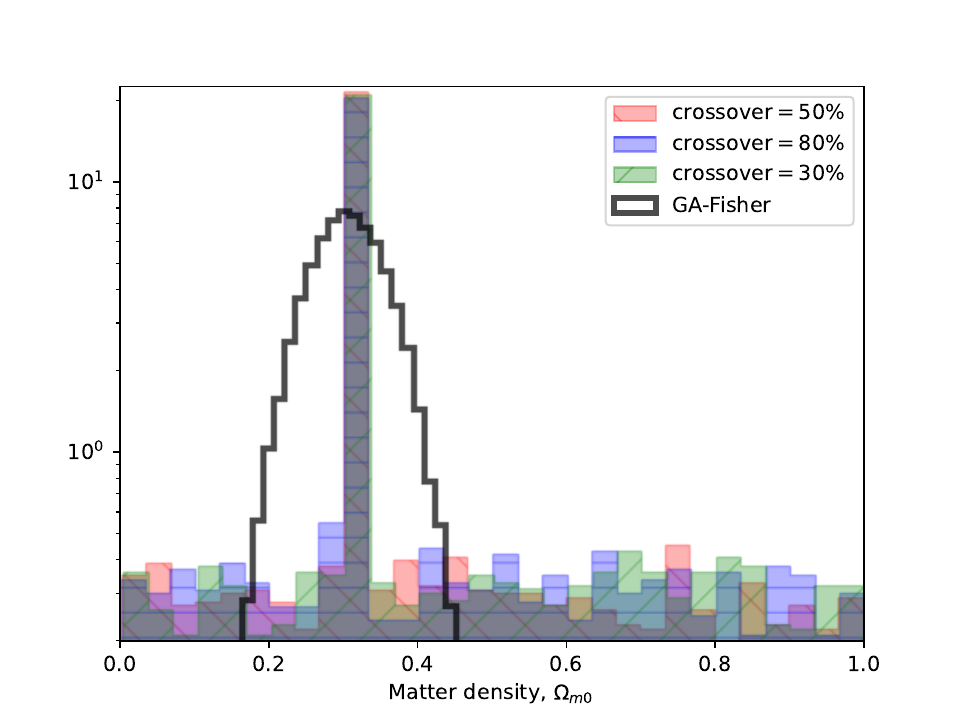}
  \includegraphics[width=0.495\textwidth]{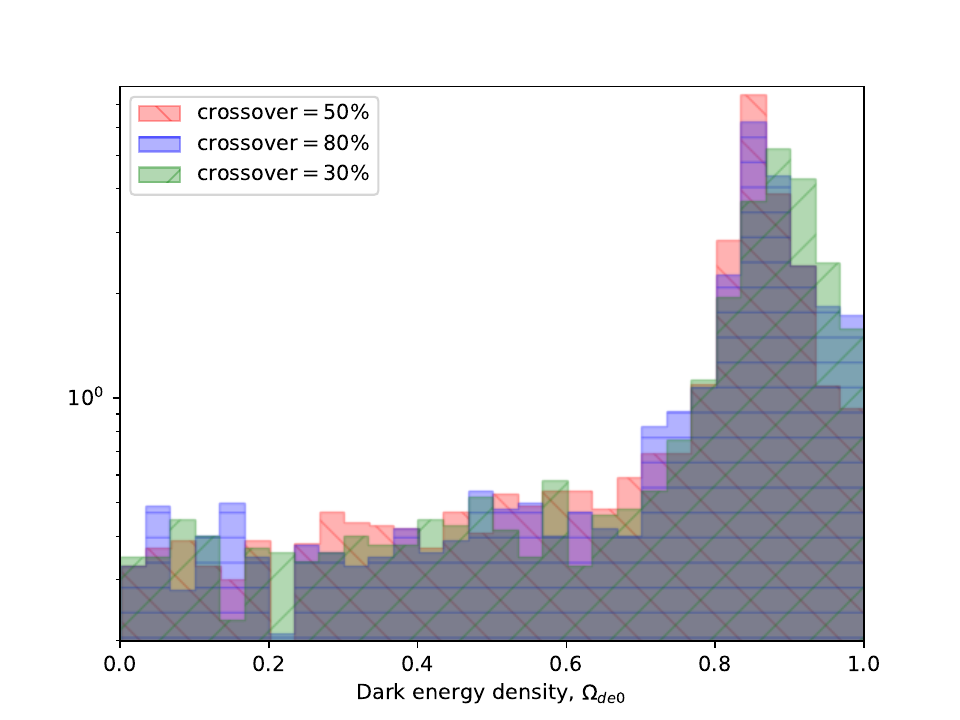}
  \includegraphics[width=0.495\textwidth]{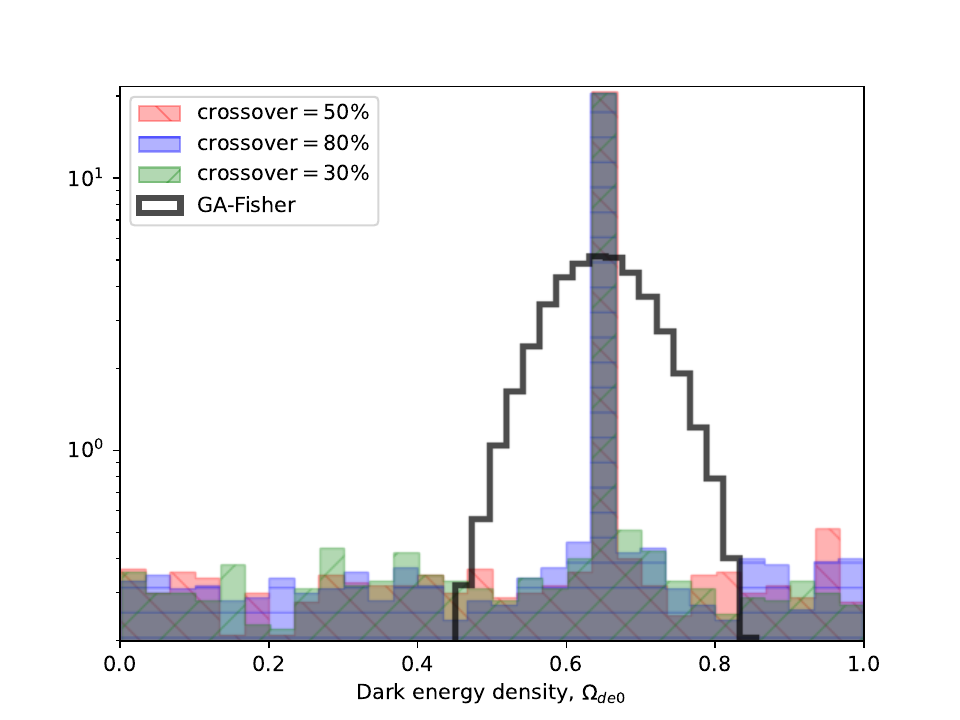}
  \caption{Changing crossover---GA final population distribution for different crossover rates; crossover=$50\%, 80\%, 30\%$ with fixed fitness function $\exp(-\chi^2/2)$ and mutation rate $(0.5, 0.3)$. {(Left column) results obtained with only CC data; (right column) results with CC and SNe.} In the right column, the Gaussian (GA-Fisher) corresponds to the GA best solution compounded with a Fisher matrix uncertainty estimate based on the likelihood.}
  \label{fig:changing_crossover}
\end{figure*}

We lastly explore the crossover rate's impact in GA. Analogous to the previous sections, we fix the fitness function to FF$_3$ \eqref{eq:ff3}, and consider a population size of 3000, a roulette wheel selection with a fixed selection rate of 30\%, and an adaptive mutation rate probability of $m_1=(50\%,30\%)$. We note that we could have similarly chosen $m_2$ as long as the mutation rate is fixed so that the results can attributed to changes in other variables. We consider three crossover scenarios, one with $c_1=50\%$, $c_2=80\%$, and $c_3=30\%$. In GA, the scattered crossover mechanism combines genes from two parents based on a randomly generated binary mask. With a $c_i$ crossover probability, there is a $c_i$ chance that the crossover will occur, meaning new offspring are produced less frequently, often resulting in offspring that closely resemble one of the parents. The results obtained for each crossover $c_{1-3}$ are shown in Figure \ref{fig:changing_crossover}. 

This shows that crossover is less of a factor compared with the fitness function and the mutation rate for the localized set of parameters in the final population. One way we can understand this is through definition; crossover is the process by which the gene pool is mixed up in a population in order to increase diversity. We find that this does not produce perceivable influences to the final population regardless of the choice. Results with CC alone and with both CC and SNe support this statement. Figure \ref{fig:changing_crossover} suggests that the choice of the crossover probability has less impact on the cosmological parameter space, i.e., the final populations for each parameter are barely distinguishable and the uncertainty would only obscure small differences, if any. Admittedly, in the top right of Figure \ref{fig:changing_crossover} the Hubble constant distribution has notably shifted by a bin for the low crossover probability. This change is irrelevant, and we view our statement on the impact of the crossover probability as a rule of thumb.

\subsection{Discussion}
\label{subsec:summary}

It is worth noting that the parameter distributions obtained from GA are highly localized and non-Gaussian, regardless of the values of the hyperparameters. We could nonetheless take it as an exercise for illustrative purposes to blindly treat the final population as if it were a posterior. In this way, we can look at the marginalized statistics of the GA samples, comparing the results with a corresponding MCMC analysis (Section \ref{subsec:mcmc}). The results are shown in Table \ref{tab:summary_stats}.

\begin{table*}[h!]
    \centering
    \renewcommand{\arraystretch}{1.2} 
    \caption{GA constraints (68\% confidence intervals) on the cosmological parameters of curved $\Lambda$CDM. $H_0$ are given in units km s$^{-1}$Mpc$^{-1}$; fitness functions FF$_i$ are given by (\ref{eq:ff1}-\ref{eq:ff3}); mutation rates are $m_1=(50\%, 30\%)$, $m_2=(80\%, 20\%)$; crossover probabilities are $c_1=50\%$, $c_2=80\%$, $c_3=30\%$.}
    \begin{tabular}{|c|c|r|c|c|c|}
    \hline
    Data set & Method & Change & $H_0$ & $\Omega_{m0}$ & $\Omega_{de0}$ \\ \hline \hline
    \multirow{10}{*}{CC} & \multirow{9}{*}{GA} & Fitness (FF$_1$) & $20.2^{+59.0}_{-17.2}$ & $0.5^{+0.3}_{-0.3}$ & $0.5^{+0.3}_{-0.3}$ \\
    & & Fitness (FF$_2$) & $64.3^{+6.8}_{-26.7}$ & $0.4^{+0.3}_{-0.2}$ & $0.6^{+0.3}_{-0.4}$ \\ 
    & & Fitness (FF$_3$) & $67.8^{+1.6}_{-23.2}$ & $0.5^{+0.1}_{-0.1}$ & $0.8^{+0.1}_{-0.4}$ \\
    & & Mutation ($m_1$) & $68.3^{+1.1}_{-19.5}$ & $0.5^{+0.1}_{-0.1}$ & $0.9^{+0.1}_{-0.4}$ \\
    & & Mutation ($m_2$) & $68.1^{+0.8}_{-15.0}$ & $0.49^{+0.04}_{-0.03}$ & $0.91^{+0.02}_{-0.36}$ \\
    & & Crossover ($c_1$) & $67.9^{+1.5}_{-22.4}$ & $0.5^{+0.1}_{-0.1}$ & $0.8^{+0.1}_{-0.4}$ \\
    & & Crossover ($c_2$) & $67.6^{+2.0}_{-24.3}$ & $0.5^{+0.1}_{-0.1}$ & $0.8^{+0.1}_{-0.4}$ \\
    & & Crossover ($c_3$) & $67.8^{+2.0}_{-23.6}$ & $0.5^{+0.1}_{-0.1}$ & $0.9^{+0.1}_{-0.4}$ \\ \cline{2-6}
    & \multirow{1}{*}{MCMC} & & $65 \pm 6$ & $0.4\pm0.2$ & $0.6\pm0.3$ \\ \hline
    \multirow{10}{*}{CC+SNe} & \multirow{9}{*}{GA} & Fitness (FF$_1$) & $0^{+65}_{-0}$ & $0.5^{+0.4}_{-0.3}$ & $0.5^{+0.3}_{-0.4}$ \\
    & & Fitness (FF$_2$) & $71.9^{+1.3}_{-23.7}$ & $0.4^{+0.3}_{-0.3}$ & $0.5^{+0.3}_{-0.3}$ \\ 
    & & Fitness (FF$_3$) & $73.4^{+0.0}_{-21.4}$ & $0.320^{+0.148}_{-0.001}$ & $0.7^{+0.0}_{-0.2}$ \\
    & & Mutation ($m_1$) & $73.4^{+0.0}_{-21.0}$ & $0.340^{+0.153}_{-0.002}$ & $0.7^{+0.0}_{-0.2}$ \\
    & & Mutation ($m_2$) & $73.3^{+0.0}_{-3.1}$ & $0.313^{+0.003}_{-0.000}$ & $0.7^{+0.0}_{-0.0}$ \\
    & & Crossover ($c_1$) & $73.2^{+0.0}_{-23.1}$ & $0.3120^{+0.1210}_{-0.0004}$ & $0.648^{+0.003}_{-0.134}$ \\
    & & Crossover ($c_2$) & $73.3^{+0.0}_{-22.6}$ & $0.3^{+0.2}_{-0.0}$ & $0.658^{+0.002}_{-0.145}$ \\
    & & Crossover ($c_3$) & $73.2^{+0.0}_{-19.9}$ & $0.3^{+0.2}_{-0.0}$ & $0.7^{+0.0}_{-0.2}$ \\ \cline{2-6}
    & \multirow{1}{*}{MCMC} & & $73.2 \pm 0.3$ & $0.3 \pm 0.1$ & $0.6\pm 0.1$ \\ \hline
    \end{tabular}
    \label{tab:summary_stats}
\end{table*}

{Before we discuss further, a few comments on convergence that are owed are in order. For MCMC or GA, we made sure that the results are reproducible given the algorithm parameters reported in the paper. In particular, in the case of MCMC, we have made sure that the number of steps or samples is sufficient to produce stable posterior distributions and uncertainty estimates to the significant figures shown in Table 1. We held the same convergence criteria for our GA applications. Alternatively, we could have applied rigorous convergence criteria such as Gelman-Rubin to MCMC \cite{doi:10.1177/096228029600500402, Gelman:1992zz} or an analogous termination condition for GA (e.g., when the fitness of a population plateaus after a large number of generations). Our results will hold, but it is always a useful exercise to test convergence (via our supplementary notebook) for readers that are unfamiliar, but want to know more about the subject.}

These show the uncertainty estimates of the cosmological parameters in GA based on the final evolved population. As we alluded to already, the shape of the distribution of samples in the final population are generally non-Gaussian; an implication of this goes to the upper and lower confidence limits as shown in Table \ref{tab:summary_stats} which are highly influenced by prior on the parameter space being investigated. This can be understood solely due to the fact that GA is not meant to converge to a posterior, unlike MCMC; GA operations (selection, crossover, mutation) have been put in place to specifically prevent convergence in order to work well for global optimization, and it does the job well. This implies that there are always going to be outliers, residing outside the localized set of samples close to the true solution, no matter how tuned evolution is made to be. These outliers explain the huge variances in GA parameter estimation, since we used directly the evolved population for uncertainty estimation.

\begin{figure}[h!]
    \centering
    \includegraphics[width=0.475\textwidth]{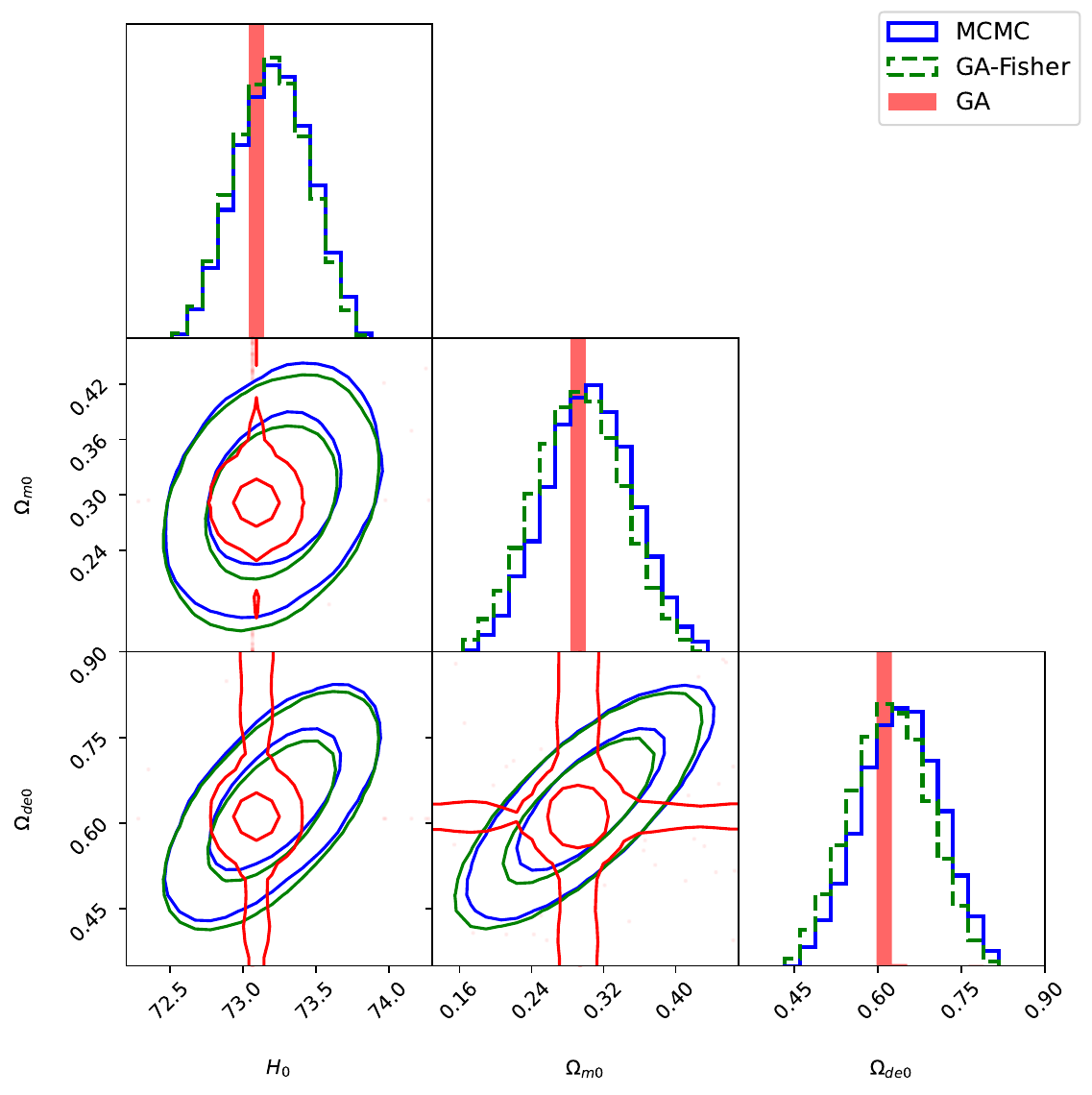}
    \caption{Constraints on curved $\Lambda$CDM with CC and SNe data obtained using MCMC and GA; MCMC (blue) results $H_0=73.21^{+0.28}_{-0.28}$ km s$^{-1}$Mpc$^{-1}$, $\Omega_{m0}=0.31^{+0.05}_{-0.05}$, $\Omega_{de0}=0.64^{+0.08}_{-0.08}$; GA-Fisher (green) results $H_0=73.24^{+0.27}_{-0.28}$ km s$^{-1}$Mpc$^{-1}$, $\Omega_{m0}=0.32^{+0.05}_{-0.05}$, $\Omega_{de0}=0.65^{+0.08}_{-0.08}$; GA (red) gives $H_0=73.3$ km s$^{-1}$Mpc$^{-1}$, $\Omega_{m0}=0.3$, $\Omega_{de0}=0.6$ (variation from the optimal solution is negligible).}
    \label{fig:mcmc_ga_lcdmconstraints}
\end{figure}

To elucidate on this point further, we investigate the combined CC and SNe data sets, using MCMC and GA, where for GA we also present the final evolved samples as well as uncertainty estimates using the Fisher matrix. Furthermore, taking in lessons learned earlier, we use an optimized hyperparameters in order to obtain a rather localized set of samples in the final population: fitness function FF$_3$, selection rate 30\%, adaptive mutation rate $m_2=(80\%)$, and crossover probability $c_1=50\%$. This catered a final population that is sharply concentrated at $H_0=73.3$ km s$^{-1}$Mpc$^{-1}$, $\Omega_{m0}=0.3$, $\Omega_{de0}=0.6$, with no perceivable variation around it at the 68\% confidence limit. The results are shown in Figure \ref{fig:mcmc_ga_lcdmconstraints}.

It is notable that the application of the Fisher matrix in conjunction with GA to provide an uncertainty band gives comparable results to MCMC \cite{Medel-Esquivel:2023nov}. This has been established in GA applications to grammatical evolution for the reconstruction of cosmological functions \cite{Nesseris:2010ep, Nesseris:2012tt}, where other methods such as path integral approach for GA uncertainty estimation has also been explored. In this case, MCMC and the GA-Fisher matrix hybrid (GA-Fisher) have almost given the same constraints on the cosmological parameters, including the correlation in the parameters. See also \cite{DiValentino:2025sru}. However, we must warn that the GA-Fisher estimate cannot always be relied upon, particularly when there is a lack of data, such is the case when we consider only the CC data. This is the reason we did not display the GA-Fisher bands in the left panels of Figures \ref{fig:changing_fitness}-\ref{fig:changing_crossover}, since the covariance estimates did not converge. This is completely understandable because the Fisher matrix uncertainty estimation is a forecasting approach. Hence, its reliability depends strongly on the amount of input it deals with. Nonetheless, the sheer size of the SNe data makes it a reliable alternative tool for cosmological parameter estimation together with GA, as shown in Figure \ref{fig:mcmc_ga_lcdmconstraints}. On the other hand, {when two different and independent methods agree on a result, then confidence on the result and the methods increases. In this case, the result are cosmological parameters, and the methods are MCMC and GA. This is}
depicted in Figure \ref{fig:mcmc_ga_lcdmconstraints} where the GA results are sharply peaked inside the MCMC, and GA-Fisher, ellipses.

{It is instructive to compare the results of GA in our work to classical fitting; such that the role of the goodness of fit is taken by the fitness function, the randomization step by the mutation operator. The analogy can be done due to the simplicity of the problem at hand, and since the results has shown that the substantial effects on the population distribution are controlled by two hyperparameters. However, what must be clear is that GA is a global optimization method. Its advantages over classical fitting will become apparent when the problem is sufficiently sophisticated (see e.g., Figure 1 of Ref. \cite{Medel-Esquivel:2023nov}). Cosmological parameter estimation in the context of our work turns out to be simple enough to only tease out the agreement, but this will not take away the features of GA that mathematically give it an edge over classical fitting. Certainly, we are looking for a cosmological problem that would cater to GAs advantages, or disadvantages compared to other methods.}

\section{Conclusions}
\label{sec:conclusions}

We highlight that GA serves as a powerful optimization strategy, with hyperparameters that can be fine-tuned to achieve desired outcomes. For the purpose of cosmological parameter estimation, our goal was to localize the results around the optimal solution, enabling traditional interpretation similar to a posterior distribution. However, mastering this tuning process requires practice. We investigated the key hyperparameters of GA---fitness, mutation, and crossover---to assess their significant impact on the final results. Our findings indicate that fitness and mutation play crucial roles in guiding the evolution toward a population that is concentrated around the optimal solution. Additionally, while we experimented with other hyperparameters such as selection rate and type, and crossover type, they did not demonstrate a substantial impact compared to the primary ones we opted to highlight in this pedagogical study. {There were no visualizable or significantly quantifiable effect that we found on the parameter distribution based on varying the selection rate and type, unlike with mutation and crossover probabilities.}

{It is worth mentioning that while our work focused on cosmological parameter estimation where MCMC is the reasonable baseline, GA has been compared to other global optimizers such as particle swarm optimization and differential evolution. See e.g., Table 3 of \cite{6735045} or Tables 4-7 of \cite{2021arXiv211210318E}. We might also mention that compared to local optimizers such as Nelder-Mead and BFGS, GA can be expected to perform relatively slower, since GA's speed and computational complexity is tied to the population size, the number of generations, and the dimensionality of the problem. The advantages of GA will nonetheless show up inevitably when dealing with, say, multimodal likelihoods and high dimensional spaces; where local optimizers tend to struggle.}

We hope that this simple application of GA clarifies the strengths and shortcomings of the method, but most essentially why it may be considered as a promising supporting tool to MCMC for cosmological analysis. {Having both GA and MCMC results will enable us to make further deductions on the parameter space than with MCMC only.}

Looking ahead, future work can revolve around  testing GA with more complex astrophysical scenarios and cosmological models beyond curved $\Lambda$CDM, further exploring the reach of biology-inspired optimization in enhancing our understanding of the cosmos.

\section*{Acknowledgements}
{The authors express their gratitude to two anonymous referees for the constructive criticism that lead to a significant improvement in the overall quality of the manuscript.} RCB is supported by an appointment to the JRG Program at the APCTP through the Science and Technology Promotion Fund and Lottery Fund of the Korean Government, and was also supported by the Korean Local Governments in Gyeongsangbuk-do Province and Pohang City. RCB acknowledges the hospitality of the Institute of Physics, Academia Sinica {in Taipei, Taiwan} that enabled the completion of this work. The initial stages of this work were done in part during the TCA Summer Student Program 2023 supported by the NCTS Taiwan.


\begin{thebibliography}{10}
\expandafter\ifx\csname url\endcsname\relax
  \def\url#1{\texttt{#1}}\fi
\expandafter\ifx\csname urlprefix\endcsname\relax\def\urlprefix{URL }\fi
\expandafter\ifx\csname href\endcsname\relax
  \def\href#1#2{#2} \def\path#1{#1}\fi

\bibitem{rudolph1994convergence}
G.~Rudolph, Convergence analysis of canonical genetic algorithms, IEEE transactions on neural networks 5~(1) (1994) 96--101.

\bibitem{Akrami:2009hp}
Y.~Akrami, P.~Scott, J.~Edsjo, J.~Conrad, L.~Bergstrom, {A Profile Likelihood Analysis of the Constrained MSSM with Genetic Algorithms}, JHEP 04 (2010) 057.
\newblock \href {http://arxiv.org/abs/0910.3950} {\path{arXiv:0910.3950}}, \href {http://dx.doi.org/10.1007/JHEP04(2010)057} {\path{doi:10.1007/JHEP04(2010)057}}.

\bibitem{Crowder:2006wh}
J.~Crowder, N.~J. Cornish, L.~Reddinger, {Darwin meets Einstein: LISA data analysis using genetic algorithms}, Phys. Rev. D 73 (2006) 063011.
\newblock \href {http://arxiv.org/abs/gr-qc/0601036} {\path{arXiv:gr-qc/0601036}}, \href {http://dx.doi.org/10.1103/PhysRevD.73.063011} {\path{doi:10.1103/PhysRevD.73.063011}}.

\bibitem{10.7551/mitpress/1090.001.0001}
J.~H. Holland, \href{https://doi.org/10.7551/mitpress/1090.001.0001}{Adaptation in Natural and Artificial Systems: An Introductory Analysis with Applications to Biology, Control, and Artificial Intelligence}, The MIT Press, 1992.
\newblock \href {http://dx.doi.org/10.7551/mitpress/1090.001.0001} {\path{doi:10.7551/mitpress/1090.001.0001}}.
\newline\urlprefix\url{https://doi.org/10.7551/mitpress/1090.001.0001}

\bibitem{Katoch2021}
S.~Katoch, S.~S. Chauhan, V.~Kumar, A review on genetic algorithm: past, present, and future, Multimedia Tools and Applications 80~(5) (2021) 8091--8126.
\newblock \href {http://dx.doi.org/10.1007/s11042-020-10139-6} {\path{doi:10.1007/s11042-020-10139-6}}.

\bibitem{Mirjalili2019}
S.~Mirjalili, Genetic Algorithm, Springer International Publishing, Cham, 2019, pp. 43--55.

\bibitem{Katsifarakis2020}
K.~L. Katsifarakis, Y.~N. Kontos, Genetic Algorithms: A Mature Bio-inspired Optimization Technique for Difficult Problems, Springer International Publishing, Cham, 2020, pp. 3--25.

\bibitem{Thompson2024}
J.~Thompson, Genetic Algorithms and Applications, Springer Nature Singapore, Singapore, 2024, pp. 981--1006.

\bibitem{Bogdanos:2009ib}
C.~Bogdanos, S.~Nesseris, {Genetic Algorithms and Supernovae Type Ia Analysis}, JCAP 05 (2009) 006.
\newblock \href {http://arxiv.org/abs/0903.2805} {\path{arXiv:0903.2805}}, \href {http://dx.doi.org/10.1088/1475-7516/2009/05/006} {\path{doi:10.1088/1475-7516/2009/05/006}}.

\bibitem{Nesseris:2010ep}
S.~Nesseris, A.~Shafieloo, {A model independent null test on the cosmological constant}, Mon. Not. Roy. Astron. Soc. 408 (2010) 1879--1885.
\newblock \href {http://arxiv.org/abs/1004.0960} {\path{arXiv:1004.0960}}, \href {http://dx.doi.org/10.1111/j.1365-2966.2010.17254.x} {\path{doi:10.1111/j.1365-2966.2010.17254.x}}.

\bibitem{Nesseris:2012tt}
S.~Nesseris, J.~Garcia-Bellido, {A new perspective on Dark Energy modeling via Genetic Algorithms}, JCAP 11 (2012) 033.
\newblock \href {http://arxiv.org/abs/1205.0364} {\path{arXiv:1205.0364}}, \href {http://dx.doi.org/10.1088/1475-7516/2012/11/033} {\path{doi:10.1088/1475-7516/2012/11/033}}.

\bibitem{Medel-Esquivel:2023nov}
R.~Medel-Esquivel, I.~G\'omez-Vargas, A.~A.~M. S\'anchez, R.~Garc\'\i{}a-Salcedo, J.~Alberto~V\'azquez, {Cosmological Parameter Estimation with Genetic Algorithms}, Universe 10~(1) (2024) 11.
\newblock \href {http://arxiv.org/abs/2311.05699} {\path{arXiv:2311.05699}}, \href {http://dx.doi.org/10.3390/universe10010011} {\path{doi:10.3390/universe10010011}}.

\bibitem{DiValentino:2025sru}
E.~Di~Valentino, et~al., {The CosmoVerse White Paper: Addressing observational tensions in cosmology with systematics and fundamental physics}\href {http://arxiv.org/abs/2504.01669} {\path{arXiv:2504.01669}}.

\bibitem{DiValentino:2020vhf}
E.~Di~Valentino, et~al., {Snowmass2021 - Letter of interest cosmology intertwined I: Perspectives for the next decade}, Astropart. Phys. 131 (2021) 102606.
\newblock \href {http://arxiv.org/abs/2008.11283} {\path{arXiv:2008.11283}}, \href {http://dx.doi.org/10.1016/j.astropartphys.2021.102606} {\path{doi:10.1016/j.astropartphys.2021.102606}}.

\bibitem{DiValentino:2020zio}
E.~Di~Valentino, et~al., {Snowmass2021 - Letter of interest cosmology intertwined II: The hubble constant tension}, Astropart. Phys. 131 (2021) 102605.
\newblock \href {http://arxiv.org/abs/2008.11284} {\path{arXiv:2008.11284}}, \href {http://dx.doi.org/10.1016/j.astropartphys.2021.102605} {\path{doi:10.1016/j.astropartphys.2021.102605}}.

\bibitem{DiValentino:2020vvd}
E.~Di~Valentino, et~al., {Cosmology intertwined III: $f\sigma_8$ and $S_8$}, Astropart. Phys. 131 (2021) 102604.
\newblock \href {http://arxiv.org/abs/2008.11285} {\path{arXiv:2008.11285}}, \href {http://dx.doi.org/10.1016/j.astropartphys.2021.102604} {\path{doi:10.1016/j.astropartphys.2021.102604}}.

\bibitem{Schoneberg:2021qvd}
N.~Sch\"oneberg, G.~Franco~Abell\'an, A.~P\'erez~S\'anchez, S.~J. Witte, V.~Poulin, J.~Lesgourgues, {The H0 Olympics: A fair ranking of proposed models}, Phys. Rept. 984 (2022) 1--55.
\newblock \href {http://arxiv.org/abs/2107.10291} {\path{arXiv:2107.10291}}, \href {http://dx.doi.org/10.1016/j.physrep.2022.07.001} {\path{doi:10.1016/j.physrep.2022.07.001}}.

\bibitem{Aghanim:2018eyx}
N.~Aghanim, et~al., {Planck 2018 results. VI. Cosmological parameters}, Astron. Astrophys. 641 (2020) A6.
\newblock \href {http://arxiv.org/abs/1807.06209} {\path{arXiv:1807.06209}}, \href {http://dx.doi.org/10.1051/0004-6361/201833910} {\path{doi:10.1051/0004-6361/201833910}}.

\bibitem{Riess:2021jrx}
A.~G. Riess, et~al., {A Comprehensive Measurement of the Local Value of the Hubble Constant with 1 km/s/Mpc Uncertainty from the Hubble Space Telescope and the SH0ES Team}, Astrophys. J. Lett. 934~(1) (2022) L7.
\newblock \href {http://arxiv.org/abs/2112.04510} {\path{arXiv:2112.04510}}, \href {http://dx.doi.org/10.3847/2041-8213/ac5c5b} {\path{doi:10.3847/2041-8213/ac5c5b}}.

\bibitem{Kammerer:2025dbi}
L.~Kammerer, D.~J. Bartlett, G.~Kronberger, H.~Desmond, P.~G. Ferreira, {syren-baryon: Analytic emulators for the impact of baryons on the matter power spectrum}\href {http://arxiv.org/abs/2506.08783} {\path{arXiv:2506.08783}}.

\bibitem{Sui:2024wob}
C.~Sui, D.~J. Bartlett, S.~Pandey, H.~Desmond, P.~G. Ferreira, B.~D. Wandelt, {SYREN-NEW: Precise formulae for the linear and nonlinear matter power spectra with massive neutrinos and dynamical dark energy}, Astron. Astrophys. 698 (2025) A1.
\newblock \href {http://arxiv.org/abs/2410.14623} {\path{arXiv:2410.14623}}, \href {http://dx.doi.org/10.1051/0004-6361/202452854} {\path{doi:10.1051/0004-6361/202452854}}.

\bibitem{Bartlett:2024jes}
D.~J. Bartlett, B.~D. Wandelt, M.~Zennaro, P.~G. Ferreira, H.~Desmond, {SYREN-HALOFIT: A fast, interpretable, high-precision formula for the {\ensuremath{\Lambda}}CDM nonlinear matter power spectrum}, Astron. Astrophys. 686 (2024) A150.
\newblock \href {http://arxiv.org/abs/2402.17492} {\path{arXiv:2402.17492}}, \href {http://dx.doi.org/10.1051/0004-6361/202449854} {\path{doi:10.1051/0004-6361/202449854}}.

\bibitem{Aizpuru:2021vhd}
A.~Aizpuru, R.~Arjona, S.~Nesseris, {Machine learning improved fits of the sound horizon at the baryon drag epoch}, Phys. Rev. D 104~(4) (2021) 043521.
\newblock \href {http://arxiv.org/abs/2106.00428} {\path{arXiv:2106.00428}}, \href {http://dx.doi.org/10.1103/PhysRevD.104.043521} {\path{doi:10.1103/PhysRevD.104.043521}}.

\bibitem{Bonici:2025ltp}
M.~Bonici, G.~D'Amico, J.~Bel, C.~Carbone, {Effort: a fast and differentiable emulator for the Effective Field Theory of the Large Scale Structure of the Universe}\href {http://arxiv.org/abs/2501.04639} {\path{arXiv:2501.04639}}.

\bibitem{Bartlett:2023cyr}
D.~J. Bartlett, L.~Kammerer, G.~Kronberger, H.~Desmond, P.~G. Ferreira, B.~D. Wandelt, B.~Burlacu, D.~Alonso, M.~Zennaro, {A precise symbolic emulator of the linear matter power spectrum}, Astron. Astrophys. 686 (2024) A209.
\newblock \href {http://arxiv.org/abs/2311.15865} {\path{arXiv:2311.15865}}, \href {http://dx.doi.org/10.1051/0004-6361/202348811} {\path{doi:10.1051/0004-6361/202348811}}.

\bibitem{Piras:2023aub}
D.~Piras, A.~Spurio~Mancini, {CosmoPower-JAX: high-dimensional Bayesian inference with differentiable cosmological emulators}\href {http://arxiv.org/abs/2305.06347} {\path{arXiv:2305.06347}}, \href {http://dx.doi.org/10.21105/astro.2305.06347} {\path{doi:10.21105/astro.2305.06347}}.

\bibitem{Campagne:2023ter}
J.-E. Campagne, F.~Lanusse, J.~Zuntz, A.~Boucaud, S.~Casas, M.~Karamanis, D.~Kirkby, D.~Lanzieri, Y.~Li, A.~Peel, {JAX-COSMO: An End-to-End Differentiable and GPU Accelerated Cosmology Library}, Open J. Astrophys. 6 (2023) 1--15.
\newblock \href {http://arxiv.org/abs/2302.05163} {\path{arXiv:2302.05163}}, \href {http://dx.doi.org/10.21105/astro.2302.05163} {\path{doi:10.21105/astro.2302.05163}}.

\bibitem{Ruiz-Zapatero:2023hdf}
J.~Ruiz-Zapatero, D.~Alonso, C.~Garc{\'\i}a-Garc{\'\i}a, A.~Nicola, A.~Mootoovaloo, J.~M. Sullivan, M.~Bonici, P.~G. Ferreira, {LimberJack.jl: auto-differentiable methods for angular power spectra analyses}\href {http://arxiv.org/abs/2310.08306} {\path{arXiv:2310.08306}}, \href {http://dx.doi.org/10.21105/astro.2310.08306} {\path{doi:10.21105/astro.2310.08306}}.

\bibitem{Bonici:2023xjk}
M.~Bonici, F.~Bianchini, J.~Ruiz-Zapatero, {Capse.jl: efficient and auto-differentiable CMB power spectra emulation}\href {http://arxiv.org/abs/2307.14339} {\path{arXiv:2307.14339}}, \href {http://dx.doi.org/10.21105/astro.2307.14339} {\path{doi:10.21105/astro.2307.14339}}.

\bibitem{Karwal:2024qpt}
T.~Karwal, Y.~Patel, A.~Bartlett, V.~Poulin, T.~L. Smith, D.~N. Pfeffer, {Procoli: Profiles of cosmological likelihoods}\href {http://arxiv.org/abs/2401.14225} {\path{arXiv:2401.14225}}.

\bibitem{Balkenhol:2024sbv}
L.~Balkenhol, C.~Trendafilova, K.~Benabed, S.~Galli, {candl: cosmic microwave background analysis with a differentiable likelihood}, Astron. Astrophys. 686 (2024) A10.
\newblock \href {http://arxiv.org/abs/2401.13433} {\path{arXiv:2401.13433}}, \href {http://dx.doi.org/10.1051/0004-6361/202449432} {\path{doi:10.1051/0004-6361/202449432}}.

\bibitem{Trotta:2008qt}
R.~Trotta, {Bayes in the sky: Bayesian inference and model selection in cosmology}, Contemp. Phys. 49 (2008) 71--104.
\newblock \href {http://arxiv.org/abs/0803.4089} {\path{arXiv:0803.4089}}, \href {http://dx.doi.org/10.1080/00107510802066753} {\path{doi:10.1080/00107510802066753}}.

\bibitem{Lewis:2019xzd}
A.~Lewis, \href{https://getdist.readthedocs.io}{{GetDist: a Python package for analysing Monte Carlo samples}}\href {http://arxiv.org/abs/1910.13970} {\path{arXiv:1910.13970}}.
\newline\urlprefix\url{https://getdist.readthedocs.io}

\bibitem{2020arXiv200505290T}
J.~{Torrado}, A.~{Lewis}, {Cobaya: Code for Bayesian Analysis of hierarchical physical models}, arXiv e-prints (2020) arXiv:2005.05290\href {http://arxiv.org/abs/2005.05290} {\path{arXiv:2005.05290}}.

\bibitem{10.1214/aoap/1034625254}
A.~Gelman, W.~R. Gilks, G.~O. Roberts, Weak convergence and optimal scaling of random walk metropolis algorithms, The Annals of Applied Probability 7~(1) (1997) 110 -- 120.
\newblock \href {http://dx.doi.org/10.1214/aoap/1034625254} {\path{doi:10.1214/aoap/1034625254}}.

\bibitem{10.1093/oso/9780198523567.003.0038}
A.~Gelman, G.~O. Roberts, W.~R. Gilks, Efficient metropolis jumping rules, in: Bayesian Statistics 5: Proceedings of the Fifth Valencia International Meeting, Oxford University Press, 1996.
\newblock \href {http://dx.doi.org/10.1093/oso/9780198523567.003.0038} {\path{doi:10.1093/oso/9780198523567.003.0038}}.

\bibitem{bj/1080222083}
H.~Haario, E.~Saksman, J.~Tamminen, {An adaptive Metropolis algorithm}, Bernoulli 7~(2) (2001) 223 -- 242.

\bibitem{2010arXiv1011.4381V}
M.~{Vihola}, {Robust adaptive Metropolis algorithm with coerced acceptance rate}, arXiv e-prints (2010) arXiv:1011.4381\href {http://arxiv.org/abs/1011.4381} {\path{arXiv:1011.4381}}, \href {http://dx.doi.org/10.48550/arXiv.1011.4381} {\path{doi:10.48550/arXiv.1011.4381}}.

\bibitem{emcee}
D.~Foreman-Mackey, D.~W. Hogg, D.~Lang, J.~Goodman, {emcee: The MCMC Hammer}, Publ. Astron. Soc. Pac. 125 (2013) 306--312.
\newblock \href {http://arxiv.org/abs/1202.3665} {\path{arXiv:1202.3665}}, \href {http://dx.doi.org/10.1086/670067} {\path{doi:10.1086/670067}}.

\bibitem{Bernardo:2021mfs}
R.~C. Bernardo, J.~Levi~Said, {Towards a model-independent reconstruction approach for late-time Hubble data}, JCAP 08 (2021) 027.
\newblock \href {http://arxiv.org/abs/2106.08688} {\path{arXiv:2106.08688}}, \href {http://dx.doi.org/10.1088/1475-7516/2021/08/027} {\path{doi:10.1088/1475-7516/2021/08/027}}.

\bibitem{Gomez-Vargas:2022bsm}
I.~G\'omez-Vargas, J.~B. Andrade, J.~A. V\'azquez, {Neural networks optimized by genetic algorithms in cosmology}, Phys. Rev. D 107~(4) (2023) 043509.
\newblock \href {http://arxiv.org/abs/2209.02685} {\path{arXiv:2209.02685}}, \href {http://dx.doi.org/10.1103/PhysRevD.107.043509} {\path{doi:10.1103/PhysRevD.107.043509}}.

\bibitem{Bainbridge2017a}
M.~B. {Bainbridge}, J.~K. {Webb}, {Artificial intelligence applied to the automatic analysis of absorption spectra. Objective measurement of the fine structure constant}, Mon. Not. Roy. Astron. Soc. 468~(2) (2017) 1639--1670.
\newblock \href {http://arxiv.org/abs/1606.07393} {\path{arXiv:1606.07393}}, \href {http://dx.doi.org/10.1093/mnras/stx179} {\path{doi:10.1093/mnras/stx179}}.

\bibitem{Bainbridge2017b}
M.~B. {Bainbridge}, J.~K. {Webb}, {Evaluating the New Automatic Method for the Analysis of Absorption Spectra Using Synthetic Spectra}, Universe 3~(2) (2017) 34.
\newblock \href {http://arxiv.org/abs/1704.08710} {\path{arXiv:1704.08710}}, \href {http://dx.doi.org/10.3390/universe3020034} {\path{doi:10.3390/universe3020034}}.

\bibitem{Lee2020AI-VPFIT}
C.-C. {Lee}, J.~K. {Webb}, R.~F. {Carswell}, D.~{Milakovi{\'c}}, {Artificial intelligence and quasar absorption system modelling; application to fundamental constants at high redshift}, Mon. Not. Roy. Astron. Soc. 504~(2) (2021) 1787--1800.
\newblock \href {http://arxiv.org/abs/2008.02583} {\path{arXiv:2008.02583}}, \href {http://dx.doi.org/10.1093/mnras/stab977} {\path{doi:10.1093/mnras/stab977}}.

\bibitem{Webb2021}
J.~K. {Webb}, C.-C. {Lee}, R.~F. {Carswell}, D.~{Milakovi{\'c}}, {Getting the model right: an information criterion for spectroscopy}, Mon. Not. Roy. Astron. Soc. 501~(2) (2021) 2268--2278.
\newblock \href {http://arxiv.org/abs/2009.08336} {\path{arXiv:2009.08336}}, \href {http://dx.doi.org/10.1093/mnras/staa3551} {\path{doi:10.1093/mnras/staa3551}}.

\bibitem{Feroz:2011bj}
F.~Feroz, K.~Cranmer, M.~Hobson, R.~Ruiz~de Austri, R.~Trotta, {Challenges of Profile Likelihood Evaluation in Multi-Dimensional SUSY Scans}, JHEP 06 (2011) 042.
\newblock \href {http://arxiv.org/abs/1101.3296} {\path{arXiv:1101.3296}}, \href {http://dx.doi.org/10.1007/JHEP06(2011)042} {\path{doi:10.1007/JHEP06(2011)042}}.

\bibitem{SCHMITT20011}
L.~M. Schmitt, Theory of genetic algorithms, Theoretical Computer Science 259~(1) (2001) 1--61.
\newblock \href {http://dx.doi.org/https://doi.org/10.1016/S0304-3975(00)00406-0} {\path{doi:https://doi.org/10.1016/S0304-3975(00)00406-0}}.

\bibitem{2011arXiv1105.3538W}
A.~H. {Wright}, {The Exact Schema Theorem}, arXiv e-prints (2011) arXiv:1105.3538\href {http://arxiv.org/abs/1105.3538} {\path{arXiv:1105.3538}}, \href {http://dx.doi.org/10.48550/arXiv.1105.3538} {\path{doi:10.48550/arXiv.1105.3538}}.

\bibitem{pygad}
A.~{Fawzy Gad}, Pygad: an intuitive genetic algorithm python library, Multimed. Tools Appl. 83 (2023) 58029--58042.
\newblock \href {http://arxiv.org/abs/2106.06158} {\path{arXiv:2106.06158}}, \href {http://dx.doi.org/10.1007/s11042-023-17167-y} {\path{doi:10.1007/s11042-023-17167-y}}.

\bibitem{Moresco:2024wmr}
M.~Moresco, {Measuring the expansion history of the Universe with cosmic chronometers}\href {http://arxiv.org/abs/2412.01994} {\path{arXiv:2412.01994}}.

\bibitem{2010JCAP...02..008S}
D.~{Stern}, R.~{Jimenez}, L.~{Verde}, M.~{Kamionkowski}, S.~A. {Stanford}, {Cosmic chronometers: constraining the equation of state of dark energy. I: H(z) measurements}, JCAP 2010~(2) (2010) 008.
\newblock \href {http://arxiv.org/abs/0907.3149} {\path{arXiv:0907.3149}}, \href {http://dx.doi.org/10.1088/1475-7516/2010/02/008} {\path{doi:10.1088/1475-7516/2010/02/008}}.

\bibitem{2012JCAP...08..006M}
M.~{Moresco}, et~al., {Improved constraints on the expansion rate of the Universe up to z \raisebox{-0.5ex}\textasciitilde 1.1 from the spectroscopic evolution of cosmic chronometers}, JCAP 2012~(8) (2012) 006.
\newblock \href {http://arxiv.org/abs/1201.3609} {\path{arXiv:1201.3609}}, \href {http://dx.doi.org/10.1088/1475-7516/2012/08/006} {\path{doi:10.1088/1475-7516/2012/08/006}}.

\bibitem{2014RAA....14.1221Z}
C.~{Zhang}, H.~{Zhang}, S.~{Yuan}, S.~{Liu}, T.-J. {Zhang}, Y.-C. {Sun}, {Four new observational H(z) data from luminous red galaxies in the Sloan Digital Sky Survey data release seven}, Research in Astronomy and Astrophysics 14~(10) (2014) 1221--1233.
\newblock \href {http://arxiv.org/abs/1207.4541} {\path{arXiv:1207.4541}}, \href {http://dx.doi.org/10.1088/1674-4527/14/10/002} {\path{doi:10.1088/1674-4527/14/10/002}}.

\bibitem{Moresco:2015cya}
M.~Moresco, {Raising the bar: new constraints on the Hubble parameter with cosmic chronometers at z \ensuremath{\sim} 2}, Mon. Not. Roy. Astron. Soc. 450~(1) (2015) L16--L20.
\newblock \href {http://arxiv.org/abs/1503.01116} {\path{arXiv:1503.01116}}, \href {http://dx.doi.org/10.1093/mnrasl/slv037} {\path{doi:10.1093/mnrasl/slv037}}.

\bibitem{Moresco:2016mzx}
M.~Moresco, L.~Pozzetti, A.~Cimatti, R.~Jimenez, C.~Maraston, L.~Verde, D.~Thomas, A.~Citro, R.~Tojeiro, D.~Wilkinson, {A 6\% measurement of the Hubble parameter at $z\sim0.45$: direct evidence of the epoch of cosmic re-acceleration}, JCAP 05 (2016) 014.
\newblock \href {http://arxiv.org/abs/1601.01701} {\path{arXiv:1601.01701}}, \href {http://dx.doi.org/10.1088/1475-7516/2016/05/014} {\path{doi:10.1088/1475-7516/2016/05/014}}.

\bibitem{Ratsimbazafy:2017vga}
A.~L. Ratsimbazafy, S.~I. Loubser, S.~M. Crawford, C.~M. Cress, B.~A. Bassett, R.~C. Nichol, P.~V\"ais\"anen, {Age-dating Luminous Red Galaxies observed with the Southern African Large Telescope}, Mon. Not. Roy. Astron. Soc. 467~(3) (2017) 3239--3254.
\newblock \href {http://arxiv.org/abs/1702.00418} {\path{arXiv:1702.00418}}, \href {http://dx.doi.org/10.1093/mnras/stx301} {\path{doi:10.1093/mnras/stx301}}.

\bibitem{Brout:2021mpj}
D.~Brout, et~al., {The Pantheon+ Analysis: SuperCal-Fragilistic Cross Calibration, Retrained SALT2 Light Curve Model, and Calibration Systematic Uncertainty}\href {http://arxiv.org/abs/2112.03864} {\path{arXiv:2112.03864}}.

\bibitem{Brout:2022vxf}
D.~Brout, et~al., {The Pantheon+ Analysis: Cosmological Constraints}, Astrophys. J. 938~(2) (2022) 110.
\newblock \href {http://arxiv.org/abs/2202.04077} {\path{arXiv:2202.04077}}, \href {http://dx.doi.org/10.3847/1538-4357/ac8e04} {\path{doi:10.3847/1538-4357/ac8e04}}.

\bibitem{Scolnic:2021amr}
D.~Scolnic, et~al., {The Pantheon+ Analysis: The Full Data Set and Light-curve Release}, Astrophys. J. 938~(2) (2022) 113.
\newblock \href {http://arxiv.org/abs/2112.03863} {\path{arXiv:2112.03863}}, \href {http://dx.doi.org/10.3847/1538-4357/ac8b7a} {\path{doi:10.3847/1538-4357/ac8b7a}}.

\bibitem{Moresco:2023zys}
M.~Moresco, {Addressing the Hubble tension with cosmic chronometers}\href {http://arxiv.org/abs/2307.09501} {\path{arXiv:2307.09501}}.

\bibitem{Moresco:2018xdr}
M.~Moresco, R.~Jimenez, L.~Verde, L.~Pozzetti, A.~Cimatti, A.~Citro, {Setting the Stage for Cosmic Chronometers. I. Assessing the Impact of Young Stellar Populations on Hubble Parameter Measurements}, Astrophys. J. 868~(2) (2018) 84.
\newblock \href {http://arxiv.org/abs/1804.05864} {\path{arXiv:1804.05864}}, \href {http://dx.doi.org/10.3847/1538-4357/aae829} {\path{doi:10.3847/1538-4357/aae829}}.

\bibitem{Moresco:2020fbm}
M.~Moresco, R.~Jimenez, L.~Verde, A.~Cimatti, L.~Pozzetti, {Setting the Stage for Cosmic Chronometers. II. Impact of Stellar Population Synthesis Models Systematics and Full Covariance Matrix}, Astrophys. J. 898~(1) (2020) 82.
\newblock \href {http://arxiv.org/abs/2003.07362} {\path{arXiv:2003.07362}}, \href {http://dx.doi.org/10.3847/1538-4357/ab9eb0} {\path{doi:10.3847/1538-4357/ab9eb0}}.

\bibitem{SupernovaCosmologyProject:1997zqe}
S.~Perlmutter, et~al., {Discovery of a supernova explosion at half the age of the Universe and its cosmological implications}, Nature 391 (1998) 51--54.
\newblock \href {http://arxiv.org/abs/astro-ph/9712212} {\path{arXiv:astro-ph/9712212}}, \href {http://dx.doi.org/10.1038/34124} {\path{doi:10.1038/34124}}.

\bibitem{SupernovaSearchTeam:1998fmf}
A.~G. Riess, et~al., {Observational evidence from supernovae for an accelerating universe and a cosmological constant}, Astron. J. 116 (1998) 1009--1038.
\newblock \href {http://arxiv.org/abs/astro-ph/9805201} {\path{arXiv:astro-ph/9805201}}, \href {http://dx.doi.org/10.1086/300499} {\path{doi:10.1086/300499}}.

\bibitem{SupernovaCosmologyProject:1998vns}
S.~Perlmutter, et~al., {Measurements of $\Omega$ and $\Lambda$ from 42 High Redshift Supernovae}, Astrophys. J. 517 (1999) 565--586.
\newblock \href {http://arxiv.org/abs/astro-ph/9812133} {\path{arXiv:astro-ph/9812133}}, \href {http://dx.doi.org/10.1086/307221} {\path{doi:10.1086/307221}}.

\bibitem{HST:2000azd}
W.~L. Freedman, et~al., {Final results from the Hubble Space Telescope key project to measure the Hubble constant}, Astrophys. J. 553 (2001) 47--72.
\newblock \href {http://arxiv.org/abs/astro-ph/0012376} {\path{arXiv:astro-ph/0012376}}, \href {http://dx.doi.org/10.1086/320638} {\path{doi:10.1086/320638}}.

\bibitem{SNLS:2005qlf}
P.~Astier, et~al., {The Supernova Legacy Survey: Measurement of $\Omega_M$, $\Omega_\Lambda$ and ${\cal w}$ from the first year data set}, Astron. Astrophys. 447 (2006) 31--48.
\newblock \href {http://arxiv.org/abs/astro-ph/0510447} {\path{arXiv:astro-ph/0510447}}, \href {http://dx.doi.org/10.1051/0004-6361:20054185} {\path{doi:10.1051/0004-6361:20054185}}.

\bibitem{Riess:2016jrr}
A.~G. Riess, et~al., {A 2.4\% Determination of the Local Value of the Hubble Constant}, Astrophys. J. 826~(1) (2016) 56.
\newblock \href {http://arxiv.org/abs/1604.01424} {\path{arXiv:1604.01424}}, \href {http://dx.doi.org/10.3847/0004-637X/826/1/56} {\path{doi:10.3847/0004-637X/826/1/56}}.

\bibitem{Pan-STARRS1:2017jku}
D.~M. Scolnic, et~al., {The Complete Light-curve Sample of Spectroscopically Confirmed SNe Ia from Pan-STARRS1 and Cosmological Constraints from the Combined Pantheon Sample}, Astrophys. J. 859~(2) (2018) 101.
\newblock \href {http://arxiv.org/abs/1710.00845} {\path{arXiv:1710.00845}}, \href {http://dx.doi.org/10.3847/1538-4357/aab9bb} {\path{doi:10.3847/1538-4357/aab9bb}}.

\bibitem{Pasten:2023rpc}
E.~Past\'en, V.~H. C\'ardenas, {Testing \ensuremath{\Lambda}CDM cosmology in a binned universe: Anomalies in the deceleration parameter}, Phys. Dark Univ. 40 (2023) 101224.
\newblock \href {http://arxiv.org/abs/2301.10740} {\path{arXiv:2301.10740}}, \href {http://dx.doi.org/10.1016/j.dark.2023.101224} {\path{doi:10.1016/j.dark.2023.101224}}.

\bibitem{Perivolaropoulos:2023iqj}
L.~Perivolaropoulos, F.~Skara, {On the homogeneity of SnIa absolute magnitude in the Pantheon+~sample}, Mon. Not. Roy. Astron. Soc. 520~(4) (2023) 5110--5125.
\newblock \href {http://arxiv.org/abs/2301.01024} {\path{arXiv:2301.01024}}, \href {http://dx.doi.org/10.1093/mnras/stad451} {\path{doi:10.1093/mnras/stad451}}.

\bibitem{Verde:2009tu}
L.~Verde, {Statistical methods in cosmology}, Lect. Notes Phys. 800 (2010) 147--177.
\newblock \href {http://arxiv.org/abs/0911.3105} {\path{arXiv:0911.3105}}, \href {http://dx.doi.org/10.1007/978-3-642-10598-2_4} {\path{doi:10.1007/978-3-642-10598-2_4}}.

\bibitem{Wolz:2012sr}
L.~Wolz, M.~Kilbinger, J.~Weller, T.~Giannantonio, {On the Validity of Cosmological Fisher Matrix Forecasts}, JCAP 09 (2012) 009.
\newblock \href {http://arxiv.org/abs/1205.3984} {\path{arXiv:1205.3984}}, \href {http://dx.doi.org/10.1088/1475-7516/2012/09/009} {\path{doi:10.1088/1475-7516/2012/09/009}}.

\bibitem{Schafer:2016vyy}
B.~M. Sch{\"a}fer, R.~Reischke, {Describing variations of the Fisher-matrix across parameter space}, Mon. Not. Roy. Astron. Soc. 460~(3) (2016) 3398--3406.
\newblock \href {http://arxiv.org/abs/1603.03626} {\path{arXiv:1603.03626}}, \href {http://dx.doi.org/10.1093/mnras/stw1221} {\path{doi:10.1093/mnras/stw1221}}.

\bibitem{Bernardo:2025pua}
R.~C. Bernardo, D.~Grand{\'o}n, J.~Levi~Said, V.~H. C{\'a}rdenas, G.~C. Belinario, R.~Reyes, {Cosmo-Learn: code for learning cosmology using different methods and mock data}\href {http://arxiv.org/abs/2508.20971} {\path{arXiv:2508.20971}}.

\bibitem{doi:10.1177/096228029600500402}
A.~Gelman, D.~B. Rubin, \href{https://doi.org/10.1177/096228029600500402}{Markov chain monte carlo methods in biostatistics}, Statistical Methods in Medical Research 5~(4) (1996) 339--355, pMID: 9004377.
\newblock \href {http://dx.doi.org/10.1177/096228029600500402} {\path{doi:10.1177/096228029600500402}}.
\newline\urlprefix\url{https://doi.org/10.1177/096228029600500402}

\bibitem{Gelman:1992zz}
A.~Gelman, D.~B. Rubin, {Inference from Iterative Simulation Using Multiple Sequences}, Statist. Sci. 7 (1992) 457--472.
\newblock \href {http://dx.doi.org/10.1214/ss/1177011136} {\path{doi:10.1214/ss/1177011136}}.

\bibitem{6735045}
S.~P. Lim, H.~Haron, Performance comparison of genetic algorithm, differential evolution and particle swarm optimization towards benchmark functions, in: 2013 IEEE Conference on Open Systems (ICOS), 2013, pp. 41--46.
\newblock \href {http://dx.doi.org/10.1109/ICOS.2013.6735045} {\path{doi:10.1109/ICOS.2013.6735045}}.

\bibitem{2021arXiv211210318E}
E.~A.~T. {Enriquez}, R.~G. {Mendoza}, A.~C.~T. {Velasco}, {Philippine Eagle Optimization Algorithm}, arXiv e-prints (2021) arXiv:2112.10318\href {http://arxiv.org/abs/2112.10318} {\path{arXiv:2112.10318}}, \href {http://dx.doi.org/10.48550/arXiv.2112.10318} {\path{doi:10.48550/arXiv.2112.10318}}.

\end{thebibliography}

\end{document}